\documentclass[showpacs,preprintnumbers,amsmath,amssymb]{revtex4}

\usepackage{graphicx}
\usepackage{dcolumn}
\usepackage{bm}

\newcommand{\bq}{\begin{equation}}
\newcommand{\eq}{\end{equation}}
\newcommand{\bqa}{\begin{eqnarray}}
\newcommand{\eqa}{\end{eqnarray}}
\newcommand{\nn}{\nonumber \\}
\newcommand{\ij}{\langle i j \rangle}

\def\be     {\begin{equation}}
\def\ee     {\end{equation}}
\def\bea        {\begin{eqnarray}}
\def\eea        {\end{eqnarray}}
\def\bnn    {\begin{eqnarray*}}
\def\enn    {\end{eqnarray*}}

\begin{document}

\title{Antiferromagnetic metal to heavy-fermion metal quantum
phase transition in the Kondo lattice model: A strong coupling
approach}
\author{Ki-Seok Kim and Mun Dae Kim}
\affiliation{ School of Physics, Korea Institute for Advanced
Study, Seoul 130-012, Korea }

\begin{abstract}
We study the quantum phase transition from an antiferromagnetic
metal to a heavy fermion metal in the Kondo lattice model. Based
on the strong coupling approach we {\it first} diagonalize the Kondo coupling term.
Since this strong coupling approach makes the
resulting Kondo term {\it relevant}, the Kondo
hybridization persists  even in the antiferromagnetic metal,
indicating that fluctuations of Kondo singlets are not critical in
the phase transition. We find that the quantum transition in our
strong coupling approach results from {\it softening of
antiferromagnetic spin fluctuations of localized spins}, driven by
the Kondo interaction. Thus, the volume change of Fermi surface
becomes continuous across the transition. Using the boson
representation of the localized spin $\vec{n}_{i}
=\frac{1}{2}z_{i\sigma}^{\dagger}\vec{\tau}_{\sigma\sigma'}z_{i\sigma'}$
with the {\it spin-fractionalized} excitation $z_{i\sigma}$, we
derive an effective U(1) gauge Lagrangian in terms of renormalized
conduction electrons and fractionalized local-spin excitations
interacting via U(1) {\it gauge fluctuations}, where the
renormalized conduction electrons are given by {\it composites} of
the conduction electrons and fractionalized spin excitations.
Performing a mean field analysis based on this effective Kondo
action, we find a mean field phase diagram as a function of
$J_{K}/D$ with various densities of conduction electrons, where
$J_{K}$ is the Kondo coupling strength and $D$ the half bandwidth
of conduction electrons. The phase diagram shows a quantum
transition, resulting from {\it condensation of the
spin-fractionalized bosons}, from an antiferromagnetic metal to a
heavy fermion metal away from half filling. We show that beyond
the mean field approximation our critical field theory
characterized by the dynamic critical exponent $z = 2$ can explain
the observed non-Fermi liquid physics such as the specific heat
coefficient $\gamma \equiv C_{v}/T \sim - \ln T$ near the quantum
critical point. Furthermore, we argue that if our scenario is
applicable, there can exist a narrow region of an anomalous
metallic phase with spin gap near the quantum critical point.
\end{abstract}


\maketitle

\section{Introduction}

Nature of non-Fermi liquid physics near quantum critical points in
heavy fermion compounds is one of the central interests in modern
condensed matter physics.\cite{Review_theory} A standard
theoretical framework is the Hertz-Moriya-Millis (HMM) theory in
the Landau-Ginzburg-Wilson framework.\cite{HMM} Using a
Hubbard-Stratonovich (HS) transformation for an appropriate
interactional channel, a local order parameter can be introduced.
Integrating out fermion degrees of freedom and expanding the
resulting logarithmic action based on the noninteracting itinerant
fermion ensemble, one obtains an effective action of the local
order parameter with dissipation that results from gapless
electrons near the Fermi surface. Based on this order parameter
action, a self-consistent mean field analysis and a
renormalization group study can be performed to find non-Fermi
liquid physics near the quantum critical point, characterized by
critical fluctuations of the local order
parameter.\cite{Nagaosa_book}

However, since the expansion in deriving HMM theory from the Fermi
liquid action is basically a weak coupling approach, the effective
theory will break down in a strong coupling limit. In addition,
the presence of gapless electrons can cause nonlocal interactions
between order parameters, making a conventional treatment
unreliable in a local effective action.\cite{Gradient_expansion}
Furthermore, HMM theory is not fully self-consistent because
feedback effects to the fermion degrees of freedom by order
parameter fluctuations are not taken into account. Actually, it is
now believed that the critical field theory of order parameter
fluctuations, that is,  HMM theory cannot explain the observed
non-Fermi liquid physics in thermal and electrical properties in
the heavy fermion metals.\cite{Review_theory}

In this study, using a strong coupling approach, we derive an
effective action in terms of both bosonic and fermionic
excitations associated with order parameter fluctuations and
gapless electrons, respectively. The weak coupling approach solves
the kinetic energy term first, and treats the interaction term
perturbatively, implying that the theory is based on the
noninteracting itinerant fermion ensemble. On the other hand, in
the  atomic limit of electrons the strong coupling approach solves
the interaction term first and then, treats the kinetic energy
term perturbatively. In the present strong coupling approach we
take into account critical order parameter fluctuations and
gapless electron excitations near the Fermi surface {\it on an
equal footing}. In other words, we incorporate both bosonic and
fermionic excitations in the effective action {\it without
integrating out the gapless fermion degrees of freedom}. Hence,
our strong coupling approach goes beyond the conventional
treatment for the quantum phase transition in interacting
itinerant electrons.

We start from an antiferromagnetic phase of localized spins in the
Kondo lattice model, where the Kondo interaction term describes
the coupling between the localized spins $\vec{n}_{i}$ and conduction
spins $\vec{s}_{i} =
\frac{1}{2}c_{i\sigma}^{\dagger}\vec{\tau}_{\sigma\sigma'}c_{i\sigma'}$.
In the strong coupling approach the Kondo
interaction term can be diagonalized by using the CP$^{1}$
representation $\vec{n}_{i}\cdot\vec{\tau} =
U_{i}\tau_{3}U_{i}^{\dagger}$ with an SU(2) matrix $U_{i} = \left(
\begin{array}{cc} z_{i\uparrow} & - z_{i\downarrow}^{\dagger} \\
z_{i\downarrow} & z_{i\uparrow}^{\dagger} \end{array} \right)$ for
the localized spins and introducing renormalized electrons
$\psi_{i\sigma}$ as composites  of the conduction electrons and
spin-fractionalized bosons, $\psi_{i\sigma} =
U^{\dagger}_{i\sigma\sigma'}c_{i\sigma'}$.\cite{CP1,Kim_Kondo1,Kim_Han}
Then, the Kondo lattice model can be rewritten in terms of the
spin-fractionalized bosons $z_{i\sigma}$ and renormalized
conduction fermions $\psi_{i\sigma}$. Integrating out the fermion
fields $\psi_{i\sigma}$ and performing the gradient expansion in
the resulting logarithmic action, this conventional strong
coupling theory leads to an effective action of the $z_{i\sigma}$
bosons, which is well known in the context of the nonlinear
$\sigma$ model.\cite{CP1} In the present study, however, we take
into account both the bosons and fermions simultaneously.

The boson representation for the localized spin instead of its
fermion representation is necessary for explaining the quantum
phase transition involving an antiferromagnetic phase. The U(1)
slave-boson representation has been used conventionally for the
study of the heavy fermion phase in the Kondo lattice problem. If
the localized spins are represented by fermions and an order
parameter (slave-boson) corresponding to the Kondo hybridization
is introduced, a critical theory can be obtained in terms of both
order parameter fluctuations and fermion excitations by
integrating out the conduction
electrons.\cite{Senthil_Kondo,Kim_Kondo2} Fermions and bosons then
interact via long range interactions mediated by slave-boson U(1)
gauge fields. This critical theory successfully explains the
non-Fermi liquid physics such as the specific heat coefficient
$\gamma = C_{v}/T \sim - \ln T$ near the quantum critical point.
However, it is  difficult to explain the observed
antiferromagnetic long range order that begins to appear at the
transition point where the heavy fermion phase disappears.
Furthermore, the order parameter in the slave-boson representation
is "hidden"; no symmetry breaking associated with lattice
translations or spin rotations is involved. Thus, the transition
driven by condensation of the hidden order parameter may not be
physical.

\section{Effective action for the Kondo lattice model}

In the Hamiltonian of the Kondo lattice model \bqa && H_{KLM} = -
t\sum_{ij\sigma}c_{i\sigma}^{\dagger}c_{j\sigma} +
J_{K}\sum_{i\sigma\sigma'}\vec{S}_{i}\cdot{c}_{i\sigma}^{\dagger}\vec{\tau}_{\sigma\sigma'}c_{i\sigma'}
, \eqa the first term describes dynamics of conduction electrons
$c_{i\sigma}$ and the second term represents antiferromagnetic
exchange couplings between conduction electrons and localized
spins ${\vec S}_{i}$, where $t$ is the hopping integral of the
conduction electrons and $J_{K}$ is the Kondo coupling strength.
Using the coherent state representation for the conduction
electrons and localized spins,\cite{Fradkin_book}  the partition function of the Kondo
lattice model in the path-integral representation can be given by
\bqa && Z =
\int{Dc_{i\sigma}}{D\vec{n}_{i}}\exp \Bigl[iS\sum_{i}
\int_{0}^{\beta}{d\tau}\int_{0}^{1}{du}{\vec
n}_{{i}}\cdot\Bigl(\frac{\partial{{\vec
n}_{{i}}}}{\partial{u}}\times\frac{\partial{{\vec
n}_{{i}}}}{\partial{\tau}}\Bigr) \nn && -
\int_{0}^{\beta}{d\tau}\Bigl(\sum_{i\sigma}c_{i\sigma}^{\dagger}(\partial_{\tau}
- \mu)c_{i\sigma} -
t\sum_{ij\sigma}c_{i\sigma}^{\dagger}c_{j\sigma} +
J_{K}S\sum_{i\sigma\sigma'}\vec{n}_{i}\cdot{c}_{i\sigma}^{\dagger}\vec{\tau}_{\sigma\sigma'}c_{i\sigma'}
\Bigr) \Bigr] , \eqa where we use $\vec{S}_{i} = S\vec{n}_{i}$
with $S = 1/2$ and $|\vec{n}_{i}| = 1$, and $\mu$ is the chemical
potential. The first term corresponds to Berry phase that comes
from the path-integral quantization of spin with an additional
parameter $u$ in a unit sphere.\cite{Fradkin_book}

In order to study the antiferromagnetic phase with collinear
ordering of localized spins, we set \bqa && \vec{n}_{i}
\rightarrow (-1)^{i}\vec{n}_{i} . \eqa Then, Eq. (2) reads \bqa &&
Z = \int{Dc_{i\sigma}}{D\vec{n}_{i}}\exp \Bigl[iS\sum_{i}(-1)^{i}
\int_{0}^{\beta}{d\tau}\int_{0}^{1}{du}{\vec
n}_{{i}}\cdot\Bigl(\frac{\partial{{\vec
n}_{{i}}}}{\partial{u}}\times\frac{\partial{{\vec
n}_{{i}}}}{\partial{\tau}}\Bigr) \nn && -
\int_{0}^{\beta}{d\tau}\Bigl(\sum_{i\sigma}c_{i\sigma}^{\dagger}(\partial_{\tau}
- \mu)c_{i\sigma} -
t\sum_{ij\sigma}c_{i\sigma}^{\dagger}c_{j\sigma} +
J_{K}S\sum_{i\sigma\sigma'}(-1)^{i}{c}_{i\sigma}^{\dagger}(\vec{n}_{i}\cdot\vec{\tau})_{\sigma\sigma'}c_{i\sigma'}
\Bigr) \Bigr] . \eqa Based on this partition function we
investigate two closely-related interesting problems; how the
antiferromagnetic ordering of local spins is affected by dynamics
of conduction electrons as varying the Kondo coupling $J_{K}$, and
how the dynamics of the conduction electrons is influenced by the
change of antiferromagnetic fluctuations of the local spins.

We apply the strong coupling approach to Eq. (4) where
we first solve the Kondo coupling term.
Although the HMM theory results in a firm and successive quantum theory
for quantum phase transitions,\cite{HMM}
the naive perturbation for the Kondo coupling term
based on the itinerant fermion ensemble do not work well in
the case of strong couplings.
Using the identity \bqa && \vec{n}_{i}\cdot{\vec
\tau} = U_{i}\tau_{3}U^{\dagger}_{i} , ~~~~~~~ U_{i} = \left(
\begin{array}{cc} z_{i\uparrow} & - z_{i\downarrow}^{\dagger} \\
z_{i\downarrow} & z_{i\uparrow}^{\dagger} \end{array} \right) ,
\eqa where $z_{i\sigma}$ is a boson field carrying a spin $\sigma$
in the SU(2) matrix field $U_{i}$ and satisfies the unimodular
constraint $\sum_{\sigma}|z_{i\sigma}|^{2} = 1$, and performing
the gauge transformation \bqa \psi_{i\sigma} =
U^{\dagger}_{i\sigma\sigma'}c_{i\sigma'} , \eqa the Kondo coupling
term,
$J_{K}S\sum_{i\sigma\sigma'}(-1)^{i}{c}_{i\sigma}^{\dagger}(\vec{n}_{i}\cdot\vec{\tau})_{\sigma\sigma'}c_{i\sigma'}$,
can be represented as
$J_{K}S\sum_{i\sigma\sigma'}(-1)^{i}\psi_{i\sigma}^{\dagger}\tau^{3}_{\sigma\sigma'}\psi_{i\sigma'}$.
As a result, the two-body Kondo interaction term is represented by a one body term.
We call $z_{i\sigma}$ and $\psi_{i\sigma}$  bosonic spinon and
fermionic chargon, respectively.

In the strong coupling approach
an antiferromagnetic spin fluctuation $\vec{n}_{i}$ carrying spin
quantum number $1$ fractionalizes into bosonic spinons
$z_{i\sigma}$ with spin $1/2$, which seems to occur
through the screening   of  conduction electrons due to the strong
Kondo interaction. The components of  $\psi_{i\sigma}$ field are
given by $\psi_{i\sigma} = \left(\begin{array}{c} \psi_{i\uparrow}
\\ \psi_{i\downarrow} \end{array} \right) = \left(\begin{array}{c}
z^{\dagger}_{i\uparrow}c_{i\uparrow}
+  z_{i\downarrow}^{\dagger}c_{i\downarrow} \\
-z_{i\downarrow}c_{i\uparrow} + z_{i\uparrow}c_{i\downarrow}
\end{array} \right)$, where $\psi_{i\uparrow}$ field represents the
usual Kondo hybridization, and  $\psi_{i\downarrow}$ field the
polarization of the bosonic spinon and the conduction electron.
The fermions $\psi_{i\sigma}$ can be considered to express Kondo
resonances. Another way to describe this fractionalization is that
the conduction electrons fractionalize into the bosonic spinons
$U_{i\sigma\sigma'}$ and the fermionic chargons $\psi_{i\sigma}$,
i.e., $c_{i\sigma} = U_{i\sigma\sigma'}\psi_{i\sigma'}$. The
resulting partition function is obtained in terms of  new field
variables $\psi_{i}$ and $U_{i}$, \bqa && Z =
\int{D\psi_{i\sigma}}DU_{i\sigma\sigma'}\delta(U_{i\sigma\sigma''}^{\dagger}U_{i\sigma''\sigma'}
- \delta_{\sigma\sigma'})\exp\Bigl[-S_{B} \nn &&
-\int_{0}^{\beta}{d\tau} \Bigl\{
\sum_{i\sigma\sigma'}\psi_{i\sigma}^{\dagger}([\partial_{\tau}-\mu]\delta_{\sigma\sigma'}
+
[U^{\dagger}_{i}\partial_{\tau}U_{i}]_{\sigma\sigma'})\psi_{i\sigma'}
\nn &&
-t\sum_{ij}\psi_{i\sigma}^{\dagger}U^{\dagger}_{i\sigma\alpha}U_{j\alpha\sigma'}\psi_{j\sigma'}
+J_{K}S\sum_{i\sigma\sigma'}(-1)^{i}\psi_{i\sigma}^{\dagger}\tau^{3}_{\sigma\sigma'}\psi_{i\sigma'}
\Bigr\}\Bigr] , \eqa where $S_{B}$ is the Berry phase action. Note
that the integration measure $\int{Dc_{i\sigma}D\vec{n}_{i}}$ in
Eq. (4) is changed into
$\int{D\psi_{i\sigma}DU_{i\sigma\sigma'}}\delta(U_{i\sigma\sigma''}^{\dagger}U_{i\sigma''\sigma'}
- \delta_{\sigma\sigma'})$ in Eq. (7). Note that the chargons
(renormalized conduction electrons) and spinons (fractionalized
local spins) are now coupled in the kinetic energy term instead of
the Kondo coupling between the conduction electrons and localized
spins in Eq. (4).

A standard way to treat this nontrivial kinetic energy term is to
integrate out the chargon fields, \bqa && Z =
\int{DU_{i\sigma\sigma'}}\delta(U_{i\sigma\sigma''}^{\dagger}U_{i\sigma''\sigma'}
- \delta_{\sigma\sigma'})\exp\Bigl[-S_{B} \nn && +\mathbf{tr}\ln
\Bigl\{ [\partial_{\tau}-\mu]\delta_{\sigma\sigma'} +
J_{K}S(-1)^{i}\tau^{3}_{\sigma\sigma'} +
[U^{\dagger}_{i}\partial_{\tau}U_{i}]_{\sigma\sigma'} -
t_{ij}U^{\dagger}_{i\sigma\alpha}U_{j\alpha\sigma'} \Bigr\}\Bigr]
, \eqa where $t_{ij} = t$ is the nearest neighbor hopping. An
effective action of the spinons can be obtained by expanding the
logarithmic term for the bosonic spinons. One important difference
from the HMM theory is that the expansion parameter is $t/J_{K}$
instead of $J_{K}/t$. However this formulation has a serious defect.
Metallic physics of the conduction electrons is
not introduced  since this treatment is valid in
the atomic limit. Actually, expanding the logarithmic term in the
expansion parameter $t/J_{K}$, the resulting effective action is
known to be the O(3) nonlinear $\sigma$ model appropriate to an
{\it insulating} antiferromagnet.\cite{Nagaosa_book} Because an
additional Berry phase term appears in the effective $\sigma$
model, the two Berry phase terms  cancel each other and the
contribution of Berry phase vanishes.\cite{Kim_Kondo1} In the
following, although we develop a formulation different from the
above standard approach, we can also exclude the Berry phase term.

An important issue of this study is how to introduce physics of
the Fermi surface of the conduction electrons in the strong
coupling approach. One possible route is to decouple the
"interacting" kinetic energy into the conventional
"noninteracting" one via the HS transformation. Unfortunately,
there is a difficulty in performing the HS transformation for the
time-derivative term in Eq. (7). We consider discrete-time steps
and rewrite the partition function as \bqa && Z \approx
\int{DU_{i\tau}^{\sigma\sigma'}D\psi_{i\tau}^{\sigma}}\exp\Bigl[ -
\Bigl\{ -
\sum_{i\tau\tau'}\psi_{i\tau}^{\dagger\sigma}U_{i\tau}^{\dagger\sigma\alpha}
U_{i\tau'}^{\alpha\sigma'}\psi_{i\tau'}^{\sigma'} -
\frac{t}{J}\sum_{ij\tau}
\psi_{i\tau}^{\dagger\sigma}U_{i\tau}^{\dagger\sigma\alpha}
U_{j\tau}^{\alpha\sigma'}\psi_{j\tau}^{\sigma'} \nn && +
\frac{J_{K}S}{J}\sum_{i\tau}(-1)^{i}\psi_{i\tau}^{\dagger\sigma}
\tau^{3}_{\sigma\sigma'} \psi_{i\tau}^{\sigma'}  - \frac{\mu}{J}
\sum_{i\tau}\psi_{i\tau}^{\dagger\sigma}\psi_{i\tau}^{\sigma}
\Bigr\}\Bigr] , \eqa where $J$ is an energy scale associated with
a time step $J = 1/\Delta\tau$. This discrete-time expression can
be reduced to the original one of Eq. (7) in the limit of
$\Delta\tau \rightarrow 0$.\cite{Discrete_time}

Performing the HS transformation for the "hopping" terms in Eq.
(9), we obtain the following expression \bqa && \exp\Bigl[-
\Bigl\{ -
\sum_{i\tau\tau'}\psi_{i\tau}^{\dagger\sigma}U_{i\tau}^{\dagger\sigma\alpha}
U_{i\tau'}^{\alpha\sigma'}\psi_{i\tau'}^{\sigma'}
- \frac{t}{J}\sum_{ij\tau}
\psi_{i\tau}^{\dagger\sigma}U_{i\tau}^{\dagger\sigma\alpha}U_{j\tau}^{\alpha\sigma'}\psi_{j\tau}^{\sigma'}
\Bigr\}\Bigr] \nn && =
\int{DF_{\mu\nu}^{\sigma\sigma'}DE_{\mu\nu}^{\sigma\sigma'}}\exp\Bigl[-
\sum_{i\tau\tau'}\Bigl\{
E_{i\tau\tau'}^{\dagger{\sigma\sigma'}}F_{i\tau\tau'}^{\sigma'\sigma}
+ H.c. -
U_{i\tau}^{{\dagger}\sigma\alpha}U_{i\tau'}^{\alpha\sigma'}F_{i\tau\tau'}^{\sigma'\sigma}
-
E_{i\tau\tau'}^{\dagger\sigma\sigma'}\psi_{i\tau'}^{\sigma'}\psi_{i\tau}^{\dagger\sigma}
- H.c. \Bigr\}\nn && - \frac{t}{J}\sum_{ij\tau}\Bigl\{
E_{ij\tau}^{\dagger{\sigma\sigma'}}F_{ij\tau}^{\sigma'\sigma} +
H.c. -
U_{i\tau}^{\dagger\sigma\alpha}U_{j\tau}^{\alpha\sigma'}F_{ij\tau}^{\sigma'\sigma}
-
E_{ij\tau}^{\dagger\sigma\sigma'}\psi_{j\tau}^{\sigma'}\psi_{i\tau}^{\sigma\dagger}
- H.c. \Bigr\}\Bigr] \nn && =
\int{DF_{\mu\nu}DE_{\mu\nu}}\exp\Bigl[-
\sum_{i\tau\tau'}\mathbf{tr}\Bigl\{
E_{i\tau\tau'}^{\dagger}F_{i\tau\tau'} + H.c. -
U_{i\tau'}F_{i\tau\tau'}U^{\dagger}_{i\tau} -
\psi_{i\tau}^{\dagger}E_{i\tau\tau'}^{\dagger}\psi_{i\tau'} - H.c.
\Bigr\}\nn && - \frac{t}{J}\sum_{ij\tau}\mathbf{tr}\Bigl\{
E_{ij\tau}^{\dagger}F_{ij\tau} + H.c. -
U_{j\tau}F_{ij\tau}U^{\dagger}_{i\tau} -
\psi_{i\tau}^{\dagger}E_{ij\tau}^{\dagger}\psi_{j\tau} - H.c.
\Bigr\}\Bigr] , \eqa where $E_{i\tau\tau'}$, $F_{i\tau\tau'}$ and
$E_{ij\tau}$, $F_{ij\tau}$ are HS matrix fields associated with
hopping of  $\psi_{i\tau}$ fermions and $z_{i\tau}$ bosons in
time and space, respectively.

We make an ansatz for the hopping matrix fields \bqa &&
E_{i\tau\tau'} \equiv E_{t}e^{ia_{i\tau\tau'}\tau_{3}} , ~~~~~
E_{ij\tau} \equiv E_{r}e^{ia_{ij\tau}\tau_{3}} , \nn &&
F_{i\tau\tau'} \equiv F_{t}e^{ia_{i\tau\tau'}\tau_{3}} , ~~~~~
F_{ij\tau} = F_{r}e^{ia_{ij\tau}\tau_{3}} , \eqa where $E_{t}$,
$F_{t}$ and $E_{r}$, $F_{r}$ are longitudinal modes (amplitudes)
of the hopping parameters, and $a_{i\tau\tau'}, a_{ij\tau}$ their
transverse modes (phase fluctuations) which are time and
spatial components of U(1) gauge fields. In fact, the transverse
modes should be represented by SU(2) gauge fields generally,
but we limit our discussion in the U(1) case for simplicity. The
hopping parameters will be determined self-consistently. Inserting
Eq. (11) into Eq. (10), and using the explicit representation [Eq.
(5)] of the SU(2) matrix $U_{i}$, we obtain an effective U(1)
gauge theory of the Kondo lattice model \bqa && Z =
\int{Dz_{i\tau\sigma}D\psi_{i\tau\sigma}}{Da_{\mu\nu}}{D\lambda_{i\tau}}{DE_{t}DF_{t}DE_{r}DF_{r}}\exp\Bigl[
- \Bigl\{ \sum_{\tau\tau'}\sum_{i}E_{t}F_{t} +
\frac{t}{J}\sum_{\tau}\sum_{ij}E_{r}F_{r} \nn && -
E_{t}\sum_{\tau\tau'}\sum_{i\sigma}
\psi_{i\tau\sigma}^{\dagger}e^{i\sigma{a}_{i\tau\tau'}}\psi_{i\tau'\sigma}
- \frac{t}{J}E_{r}\sum_{\tau}\sum_{ij\sigma}
\psi_{i\tau\sigma}^{\dagger}e^{i\sigma{a}_{ij\tau}}\psi_{j\tau\sigma}
+
\frac{J_{K}S}{J}\sum_{\tau}\sum_{i\sigma}(-1)^{i}\sigma\psi_{i\tau\sigma}^{\dagger}\psi_{i
\tau\sigma}
\nn &&
- \frac{\mu}{J}
\sum_{\tau}\sum_{i\sigma}\psi_{i\tau\sigma}^{\dagger}\psi_{i\tau\sigma}
-
F_{t}\sum_{\tau\tau'}\sum_{i\sigma}z_{i\tau\sigma}^{\dagger}e^{ia_{i\tau\tau'}}z_{i\tau'\sigma}
-
\frac{t}{J}F_{r}\sum_{\tau}\sum_{ij\sigma}z_{i\tau\sigma}^{\dagger}e^{ia_{ij\tau}}z_{j\tau\sigma}
\nn && +
i\sum_{\tau}\sum_{i}\frac{\lambda_{i\tau}}{J}(\sum_{\sigma}|z_{i\tau\sigma}|^{2}
- 1)  \Bigr\}\Bigr] , \eqa where the $\tau_{3}$ matrix is replaced
with $\sigma = \pm$. $\lambda_{i}$ is a Lagrange multiplier field
to impose the unimodular constraint.

The last step in deriving an effective U(1) gauge theory of the
Kondo lattice model is to perform the limit of $\Delta \tau
\rightarrow 0$. Ignoring the time component $a_{i\tau\tau'}$ of
the U(1) gauge field for the time being, the time-derivative term
of the bosonic spinons becomes in the tight-binding approximation
for the discrete time \bqa && -
F_{t}\sum_{\tau\tau'}\sum_{i\sigma}z_{i\tau\sigma}^{\dagger}z_{i\tau'\sigma}
= -
2F_{t}\sum_{\Omega_{n}}\sum_{i\sigma}\cos(\frac{\Omega_{n}}{J})z_{i\sigma}^{\dagger}(\Omega_{n})
z_{i\sigma}(\Omega_{n}) \nn && \approx -
2F_{t}\sum_{\Omega_{n}}\sum_{i\sigma}\Bigl(1 -
\frac{1}{2}(\frac{\Omega_{n}}{J})^2\Bigr)z_{i\sigma}^{\dagger}(\Omega_{n})
z_{i\sigma}(\Omega_{n}) \nn && = \int_{0}^{\beta}{d\tau}
\sum_{i\sigma}\Bigl(\frac{F_{t}}{J}|\partial_{\tau}z_{i\sigma}|^{2}
- 2JF_{t}|z_{i\sigma}|^{2} \Bigr) . \eqa The last term can be
absorbed into the Lagrange multiplier term, thus has no physical
effects. The derivation of Eq. (13) is based on the relativistic
invariance for the bosonic spinons. This  is reasonable because
spin excitations in the antiferromagntic phase have the
$\omega_{n} \sim k$ dispersion, exhibiting the relativistic
invariance. On the other hand, the relativistic assumption for the
bosonic spinons is not appropriate for the fermionic chargons
because the chargons have Fermi surface. In this case it is
natural to perform the $\Delta \tau \rightarrow 0$ limit naively.
Then the time-derivative term for the fermionic chargons is given
by \bqa && - E_{t}\sum_{\tau\tau'}\sum_{i\sigma}
\psi_{i\tau\sigma}^{\dagger} \psi_{i\tau'\sigma} = -
E_{t}\sum_{\tau}\sum_{i\sigma}
\psi_{i\tau\sigma}^{\dagger}\Bigl(\psi_{i\tau\sigma} -
\frac{\Delta\psi_{i\tau\sigma}}{\Delta\tau}\Delta\tau\Bigr) \nn &&
=
\int_{0}^{\beta}{d\tau}\sum_{i\sigma}\Bigl(E_{t}\psi_{i\sigma}^{\dagger}\partial_{\tau}\psi_{i\sigma}
- JE_{t}\psi_{i\sigma}^{\dagger}\psi_{i\sigma} \Bigr) , \eqa where
$\Delta\tau = 1/J$ is used. The last term is also absorbed into
the chemical potential term.

Based on the above discussion, we find the effective U(1) gauge
action for the Kondo lattice model \bqa && Z =
\int{Dz_{i\sigma}D\psi_{i\sigma}}{Da_{\mu\nu}}{D\lambda_{i}}{DE_{t}DF_{t}DE_{r}DF_{r}}e^{
- S_{eff}} , \nn && S_{eff} = S_{0} + S_{\psi} + S_{z} , \nn &&
S_{0} = \int_{0}^{\beta}{d\tau}\Bigl( J\sum_{i}E_{t}F_{t} +
t\sum_{ij}E_{r}F_{r} \Bigr) , \nn && S_{\psi} =
\int_{0}^{\beta}{d\tau}\Bigl(
E_{t}\sum_{i\sigma}\psi_{i\sigma}^{\dagger}(\partial_{\tau} -
i\sigma{a}_{i\tau})\psi_{i\sigma} -
tE_{r}\sum_{ij\sigma}\psi_{i\sigma}^{\dagger}e^{i\sigma{a}_{ij}}\psi_{j\sigma}
+
J_{K}S\sum_{i\sigma}(-1)^{i}\sigma\psi_{i\sigma}^{\dagger}\psi_{i\sigma}
\nn && ~~~~~~~ - \mu
\sum_{i\sigma}\psi_{i\sigma}^{\dagger}\psi_{i\sigma} \Bigr) , \nn
&& S_{z} = \int_{0}^{\beta}{d\tau}\Bigl(
\frac{F_{t}}{J}\sum_{i\sigma}|(\partial_{\tau} -
ia_{i\tau})z_{i\sigma}|^{2} -
tF_{r}\sum_{ij\sigma}z_{i\sigma}^{\dagger}e^{ia_{ij}}z_{j\sigma} +
i\sum_{i}\lambda_{i}(\sum_{\sigma}|z_{i\sigma}|^{2} - 1) \Bigr) .
\eqa Here, both the conduction electrons and localized  spins are
taken into account on an equal footing {\it in the strong coupling
regime}, and the bosonic effective action $S_{z}$ associated with
order parameter fluctuations is derived without integrating out
the gapless conduction electrons {\it explicitly}. The fermionic
effective action $S_{\psi}$ has essentially the same structure as
the action of the conduction electrons in Eq. (4), considering
that antiferromagnetic spin fluctuations are frozen to be
$\vec{n}_{i}\cdot\vec{\tau} = \tau_{3}$ in Eq. (4), and gauge
fluctuations are ignored in Eq. (15). When the spinons are
condensed, the effective action in Eq. (15) is reduced to Eq. (4)
with $\vec{n}_{i}\cdot\vec{\tau} = \tau_{3}$. In the condensed
phase gauge fluctuations can be ignored in the low energy limit
because they are gapped due to the Anderson-Higgs mechanism. The
spinon action $S_{z}$ is equivalent to the CP$^{1}$ action of the
O(3) nonlinear $\sigma$ model.\cite{CP1} Hence, the effective
action of Eq. (15) can recover the insulating antiferromagnet at
half filling of the conduction electrons with Fermi-nesting.

\section{Mean field analysis}

Now we perform the saddle-point analysis to obtain possible phases
of the Kondo lattice model. The mean field phases will be
determined by condensation of the bosonic spinons. We ignore the
U(1) gauge fluctuations in the mean field analysis. Later, these
gauge excitations will be allowed beyond the mean field
approximation. Then, we consider the effective mean field action
\bqa && Z_{MF} = \int{Dz_{i\sigma}D\psi_{i\sigma}}\exp\Bigl[-
\Bigl\{ \int_{0}^{\beta}{d\tau}\Bigl(
E_{t}\sum_{i\sigma}\psi_{i\sigma}^{\dagger}\partial_{\tau}\psi_{i\sigma}
- tE_{r}\sum_{ij\sigma}\psi_{i\sigma}^{\dagger}\psi_{j\sigma} \nn
&&  +
J_{K}S\sum_{i\sigma}(-1)^{i}\sigma\psi_{i\sigma}^{\dagger}\psi_{i\sigma}
- \mu \sum_{i\sigma}\psi_{i\sigma}^{\dagger}\psi_{i\sigma} +
\frac{F_{t}}{J}\sum_{i\sigma}|\partial_{\tau}z_{i\sigma}|^{2} -
tF_{r}\sum_{ij\sigma}z_{i\sigma}^{\dagger}z_{j\sigma} \nn && +
\lambda\sum_{i}(\sum_{\sigma}|z_{i\sigma}|^{2} - 1) +
J\sum_{i}E_{t}F_{t} + t\sum_{ij}E_{r}F_{r} \Bigr) \Bigr\}\Bigr]
\eqa with  $\lambda \equiv i\lambda_{i}$. The hopping parameters
$E_{t}$, $F_{t}$, $E_{r}$, $F_{r}$ and the effective chemical
potentials $\lambda$, $\mu$ can be written from the saddle-point
equations as \bqa && JE_{t} =
\frac{1}{J}\sum_{\sigma}\langle|\partial_{\tau}z_{i\sigma}|^{2}\rangle
, ~~~~~~ JF_{t} =
\sum_{\sigma}\langle{\psi}_{i\sigma}^{\dagger}\partial_{\tau}\psi_{i\sigma}\rangle
, \nn && E_{r} =
\sum_{\sigma}\langle{z}_{i\sigma}^{\dagger}z_{j\sigma}\rangle ,
~~~~~~ F_{r} =
\sum_{\sigma}\langle{\psi}_{i\sigma}^{\dagger}\psi_{j\sigma}\rangle
, \nn && 1 =
\sum_{\sigma}\langle{z}_{i\sigma}^{\dagger}z_{i\sigma}\rangle  ,
~~~~~~ 1 - \delta =
\sum_{\sigma}\langle{c}_{i\sigma}^{\dagger}c_{i\sigma}\rangle =
\sum_{\sigma}\langle{\psi}_{i\sigma}^{\dagger}\psi_{i\sigma}\rangle,
 \eqa
where $\delta$ is hole concentration.

Integrating out fermions $\psi_{i\sigma}$ and bosons
$z_{i\sigma}$ in Eq. (16), we obtain the following expression for
the free energy \bqa && F_{MF} = -
\frac{1}{\beta}\sum_{\omega_{n}}\sum_{k\sigma}\mathbf{tr}\ln\left(
\begin{array}{cc} iE_{t}\omega_{n} - \mu + E_{r}\epsilon_{k}^{\psi} & \sigma{J}_{K}S \\
\sigma{J}_{K}S & iE_{t}\omega_{n} - \mu +
E_{r}\epsilon_{k+Q}^{\psi}
\end{array} \right)
\nn && + \frac{1}{\beta} \sum_{\Omega_{n}}\sum_{k\sigma}
\ln\Bigl(\frac{F_{t}}{J}\Omega_{n}^{2} + F_{r}\epsilon_{k}^{z} +
\lambda \Bigr)  + \sum_{k}(JE_{t}F_{t} + DE_{r}F_{r}) +
\sum_{k}(\mu[1-\delta] - \lambda) . \eqa Here
$\epsilon_{k}^{\psi}$ and $\epsilon_{k}^{z}$ are the bare
dispersions of chargons and spinons, respectively. $\omega_{n}$
($\Omega_{n}$) is the fermionic (bosonic) Matsubara frequency.
Minimizing the free energy in Eq. (18) with respect to $E_{t}$,
$F_{t}$, $E_{r}$, $F_{r}$, $\lambda$, and $\mu$, we obtain the
self-consistent mean field equations \bqa && JE_{t} = -
\int_{-D}^{D}d\epsilon{D}(\epsilon)\frac{2}{\beta}\sum_{\Omega_{n}}\frac{\Omega_{n}^{2}/J}{\frac{F_{t}}{J}\Omega_{n}^{2}
+F_{r}\epsilon + \lambda} ,\nn && JF_{t} =
\sum'_{k}\frac{2}{\beta}\sum_{\omega_{n}}\frac{i\omega_{n}[(iE_{t}\omega_{n}-\mu
+E_{r}\epsilon_{k+Q}^{\psi}) + (iE_{t}\omega_{n}-\mu
+E_{r}\epsilon_{k}^{\psi})]}{(iE_{t}\omega_{n}-\mu
+E_{r}\epsilon_{k}^{\psi}) (iE_{t}\omega_{n}-\mu
+E_{r}\epsilon_{k+Q}^{\psi}) - (J_{K}S)^{2}} , \nn && DE_{r} = -
\int_{-D}^{D}d\epsilon{D}(\epsilon)\frac{2}{\beta}\sum_{\Omega_{n}}\frac{\epsilon}{\frac{F_{t}}{J}\Omega_{n}^{2}
+F_{r}\epsilon + \lambda}  , \nn && DF_{r} =
\sum'_{k}\frac{2}{\beta}\sum_{\omega_{n}}\frac{(iE_{t}\omega_{n}-\mu)(\epsilon_{k}^{\psi}
+ \epsilon_{k+Q}^{\psi}) +
2E_{r}\epsilon_{k+Q}^{\psi}\epsilon_{k}^{\psi}}{(iE_{t}\omega_{n}-\mu+E_{r}\epsilon_{k}^{\psi})
(iE_{t}\omega_{n}-\mu+E_{r}\epsilon_{k+Q}^{\psi}) - (J_K{S})^{2}}
, \nn && 1 =
\int_{-D}^{D}d\epsilon{D}(\epsilon)\frac{2}{\beta}\sum_{\Omega_{n}}\frac{1}{\frac{F_{t}}{J}\Omega_{n}^{2}
+F_{r}\epsilon + \lambda} , \nn && 1 - \delta = -
\sum'_{k}\frac{2}{\beta}\sum_{\omega_{n}}\frac{[(iE_{t}\omega_{n}-\mu+E_{r}\epsilon_{k+Q}^{\psi})
+
(iE_{t}\omega_{n}-\mu+E_{r}\epsilon_{k}^{\psi})]}{(iE_{t}\omega_{n}-\mu+E_{r}\epsilon_{k}^{\psi})
(iE_{t}\omega_{n}-\mu+E_{r}\epsilon_{k+Q}^{\psi}) - (J_{K}S)^{2}}
.  \eqa Here $\sum_{k}$ is replaced with
$\int_{-D}^{D}d\epsilon{D}(\epsilon)$ in the bosonic equations,
where $D(\epsilon)$ is the density of states for the bosonic
spectrum $\epsilon_{k}^{z}$. $\sum'_{k}$ in the fermionic
equations means sum over the folded Brillouin zone. The factor $2$
in the $1/\beta$ terms comes from the spin degeneracy. The chargon
spectrum $\epsilon_{k}^{\psi}$ is given by the electron bare
dispersion in the tight binding approximation, i.e.,
$\epsilon_{k}^{\psi} = -2t(\cos{k}_{x} + \cos{k}_{y})$.
Furthermore, a constant density of states $D(\epsilon) = 1/2D$
will be used, where $D = 4t$ is a half of the band width.

In Eq. (19) we should introduce energy cutoff in the frequency
integrals for $E_{t}$ and $F_{t}$. Note that the usual momentum
cut-off $D$ is introduced in the integrals for $E_{r}$ and $F_{r}$
to prevent the divergence. Quite similarly, we also introduce an
energy cutoff $J$ in the integrals for $E_{t}$ and $F_{t}$ because
it corresponds to inverse of lattice spacing in the frequency
space. When evaluating the frequency integrals for $E_{t}$ and
$F_{t}$, we first divide the integrals into two parts, divergent
and divergent-free parts. We calculate the divergent parts within
the energy cutoff, but for the divergent-free integrals we perform
the Matsubara summation without the energy cutoff.

\subsection{Kondo insulator}

We solve the mean field equations [Eq. (19)] at half filling of
the conduction electrons, where the chargon chemical potential
$\mu$ is zero due to the particle-hole symmetry.
The Fermi-nesting induces a gap
corresponding to the Kondo coupling in the chargon spectrum.
The excitation spectrum of the chargons is
given by $E_{k}^{\psi} =
\sqrt{(E_{r}\epsilon_{k}^{\psi})^{2}+(J_{K}S)^{2}}$. Performing
the Matsubara summation in the frequency integrals in Eq. (19), we
obtain \bqa && JE_{t} = - \frac{2J}{\pi{F}_{t}} +
\frac{1}{J}\int_{-D}^{D}d\epsilon{D}(\epsilon)
\frac{\sqrt{F_{r}\epsilon+\lambda}}{(\frac{F_{t}}{J})^{3/2}}\coth(\frac{\beta}{2}\sqrt{\frac{F_{r}\epsilon+\lambda}{\frac{F_{t}}{J}}})
\nn && ~~~~~
=_{(T \rightarrow 0)} - \frac{2J}{\pi{F}_{t}} +
\frac{1}{2DJ}\int_{-D}^{D}d\epsilon
\frac{\sqrt{F_{r}\epsilon+\lambda}}{(\frac{F_{t}}{J})^{3/2}} , \nn
&& JF_{t} = \frac{2J}{\pi{E}_{t}} + \frac{2}{E_{t}^{2}}\sum'_{k}
E_{k}^{\psi}\tanh(\frac{\beta{E}_{k}^{\psi}}{2E_{t}}) =_{(T
\rightarrow 0)} \frac{2J}{\pi{E}_{t}} + \frac{1}{2DE_{t}^{2}}
\int_{-D}^{D}d\epsilon\sqrt{E_{r}^2\epsilon^{2}+(J_KS)^2} , \nn &&
DE_{r} = -
\int_{-D}^{D}d\epsilon{D}(\epsilon)\frac{\epsilon}{\frac{F_{t}}{J}\sqrt{\frac{F_{r}\epsilon+\lambda}{\frac{F_{t}}{J}}}}
\coth(\frac{\beta}{2}\sqrt{\frac{F_{r}\epsilon+\lambda}{\frac{F_{t}}{J}}})
=_{(T \rightarrow 0)} - \frac{1}{2D} \int_{-D}^{D}d\epsilon
\frac{\epsilon}{\frac{F_{t}}{J}\sqrt{\frac{F_{r}\epsilon+\lambda}{\frac{F_{t}}{J}}}}
, \nn && DF_{r} =
2E_{r}\sum'_{k}\epsilon_{k}^{\psi{2}}\frac{\tanh(\frac{\beta{E}_{k}^{\psi}}{2E_{t}})}{E_{t}E_{k}^{\psi}}
=_{(T \rightarrow 0)} \frac{E_{r}}{2D}\int_{-D}^{D}d\epsilon
\frac{\epsilon^{2}}{E_{t}\sqrt{E_{r}^2\epsilon^{2} + (J_KS)^2}} ,
\nn && 1 =
\int_{-D}^{D}d\epsilon{D}(\epsilon)\frac{1}{\frac{F_{t}}{J}\sqrt{\frac{F_{r}\epsilon+\lambda}{\frac{F_{t}}{J}}}}
\coth(\frac{\beta}{2}\sqrt{\frac{F_{r}\epsilon+\lambda}{\frac{F_{t}}{J}}})
=_{(T \rightarrow 0)} \frac{1}{2D}\int_{-D}^{D}d\epsilon
\frac{1}{\frac{F_{t}}{J}\sqrt{\frac{F_{r}\epsilon+\lambda}{\frac{F_{t}}{J}}}}
, \eqa where we use a constant density of states, $D(\epsilon) = 1/2D$.
The first terms in the equations
for $E_{t}$ and $F_{t}$ come from the divergent parts.

Performing the momentum integrals in Eq. (20), we obtain the
following expressions for the self-consistent mean field equations
of $E_{t}$, $F_{t}$, $E_{r}$, $F_{r}$, and $\lambda$ \bqa &&
JE_{t} = - \frac{2J}{\pi{F}_{t}} +
\frac{(J[\lambda+DF_{r}])^{3/2}-(J[\lambda-DF_{r}])^{3/2}}{3DJF_{r}F_{t}^{3/2}}
,\nn && JF_{t} = \frac{2J}{\pi{E}_{t}} +
\frac{DE_{r}\sqrt{D^2E_{r}^{2} + (J_KS)^2} +
(J_KS)^2\sinh^{-1}(DE_{r}/J_KS)}{2DE_{t}^{2}E_{r}} , \nn && DE_{r}
=
\frac{(2\lambda-DF_{r})\sqrt{J(\lambda+DF_{r})}-(2\lambda+DF_{r})\sqrt{J(\lambda-DF_{r})}}{3DF_{r}^{2}\sqrt{F_{t}}}
, \nn && DF_{r} =
\frac{DE_{r}\sqrt{D^2E_{r}^{2}+(J_KS)^2}-(J_KS)^2\sinh^{-1}(DE_{r}/J_KS)}{2DE_{t}E_{r}^{2}},
\nn && 1 =
\frac{\sqrt{J(\lambda+DF_{r})}-\sqrt{J(\lambda-DF_{r})}}{DF_{r}\sqrt{F_{t}}}
. \eqa Although it is not easy to obtain analytic expressions for
the mean field parameters as a function of $J_{K}/D$ and $J/D$, we
can find the quantum critical point where the bosonic spinons
begin to be condensed. The spinon condensation occurs at
$\lambda_{c} = DF_{rc}$, where $c$ denotes "critical." Inserting this
condition into Eq. (21), we find the quantum critical point defined
by \bqa && F_{tc}^{3}\sqrt{1+(\frac{3J_{Kc}S}{D})^{2}} =
\frac{24}{\pi^2}\frac{J}{D}F_{tc}^{2} -
\frac{28}{\pi}\frac{J}{D}F_{tc}+8\frac{J}{D} , ~~~~~ F_{rc} =
\frac{2}{F_{tc}}\frac{J}{D} , ~~~~~ E_{tc} = -
\frac{2}{\pi{F}_{tc}} + \frac{4}{3F_{tc}^{2}} , \nn && E_{rc} =
\frac{1}{3} , ~~~~~
F_{tc}^{3}\Bigl(\frac{3J_{Kc}S}{D}\Bigr)^{2}\sinh^{-1}\Bigl(\frac{1}{3J_{Kc}S/D}\Bigr)
= \frac{24}{\pi^2}\frac{J}{D}F_{tc}^{2} -
\frac{20}{\pi}\frac{J}{D}F_{tc} + \frac{8}{3}\frac{J}{D} . \eqa

Solving Eq. (22), we find that the quantum critical point
$J_{Kc}S/D$ depends on the value of $J/D$. In the case of $J/D =
0.01$ we obtain $J_{Kc}S/D \approx 0.09 > 0$ that completely
coincides with the result obtained by solving Eq. (21) numerically
as shown in Fig. 1. Decreasing $J/D$ from this value, the critical
value becomes larger. In the case of $J/D = 0.1$ we find
$J_{Kc}S/D \approx - 0.10 < 0$, indicating that there is no phase
transition at half filling, and only the Kondo insulating phase
(see below) appears. Increasing $J/D$ further, the critical value
gets more negative. The condition for this quantum transition can
be obtained from the boson sector of the mean field action in Eq.
(16) {\it without any detailed calculations}. Since the boson
Lagrangian coincides with the rotor model, more precisely, the
CP$^{1}$ Lagrangian of the O(3) nonlinear $\sigma$ model, one can
read the transition condition $(D/J)F_{t}F_{r} \approx 1$ from the
boson Lagrangian itself. Actually, it can be seen from Eq. (22)
that the mean field condition for the spinon condensation given by
$\lambda_{c} = DF_{rc}$ coincides with $(D/J)F_{tc}F_{rc} = 2$
exactly.

The mean field parameters in this effective Kondo action at half
filling is shown in Fig. 1, where $J/D=0.01$. For $J_K < J_{Kc}$
the condensation of  bosonic spinons occurs, indicating the
emergence of an antiferromagnetic order for the localized spins.
On the other hand, for $J_{K} > J_{Kc}$ the spinons become gapped,
implying that the localized spins are disordered and the
antiferromagnetic order vanishes. Increasing the Kondo coupling
strength, the localized spins are strongly affected by the
conduction electrons. Thus the effective hopping parameters
$F_{t}$ and $F_{r}$ decrease as increases $J_{K}$ (Fig. 1). As
$F_{t}$ and $F_{r}$ decrease further so that $(D/J)F_{t}F_{r} \ll
1$ for large Kondo couplings, quantum fluctuations of spinons get
stronger, destructing the antiferromagnetic long range order of
the localized spins.

\begin{figure}[t]
\vspace*{10cm} \includegraphics{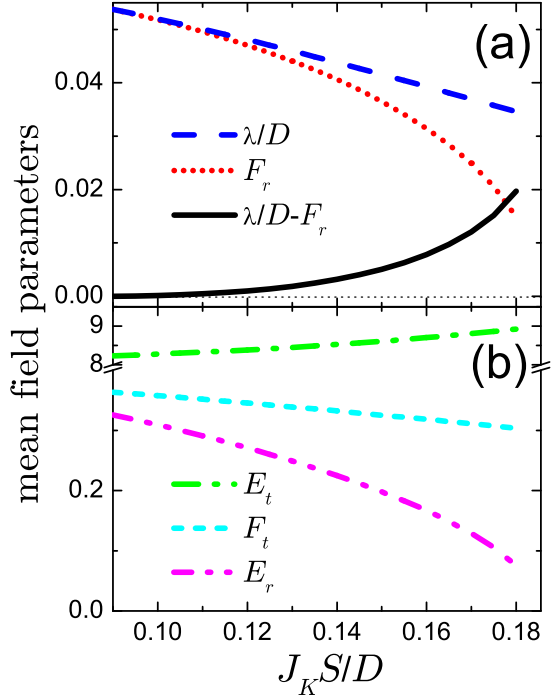} \vspace*{0cm} \caption{(Color
online) The mean field parameters at half filling are shown as
varying $J_{K}S/D$ with $J/D = 0. 01$. The quantum critical point
defined by $\lambda_{c}/D - F_{rc} = 0$ in (a) coincides exactly
with the analytic result $J_{Kc}S/D \approx 0. 09$. The effective hopping parameters $F_{r}$
and $E_{r}$ become zero at a certain value of the Kondo coupling
beyond its critical point, which implies that our spin
decomposition cannot cover the whole range of the Kondo lattice
model.}
\end{figure}

Meanwhile the Kondo hybridization
$\langle\vec{n}_{i}\cdot({c}_{i\sigma}^{\dagger}\vec{\tau}_{\sigma\sigma'}c_{i\sigma'})\rangle$
is nonzero in both phases because \bqa &&
\langle{c}_{i\sigma}^{\dagger}(\vec{n}_{i}\cdot\vec{\tau})_{\sigma\sigma'}c_{i\sigma'}\rangle
=
\langle{c}_{i\sigma}^{\dagger}U_{i\sigma\alpha}{\tau}_{3\alpha\beta}U_{i\beta\sigma'}^{\dagger}c_{i\sigma'}\rangle
=
\langle{\psi}_{i\alpha}^{\dagger}{\tau}_{3\alpha\beta}\psi_{i\beta}\rangle
= \langle{\psi}_{i\uparrow}^{\dagger}\psi_{i\uparrow} -
\psi_{i\downarrow}^{\dagger}\psi_{i\downarrow}\rangle \not= 0 ,
\eqa which is the hallmark of the present strong coupling
approach. In a different angle one may view this as an assumption
in our strong coupling theory. At half filling, due to the
Fermi-nesting the chargon excitations are gapped, thus both phases
are insulators.

When the localized spins form an antiferromagnetic order in $J_{K}
< J_{Kc}$ with  the condensation of bosonic spinons, $\langle
z_{i\sigma}\rangle\neq 0$, the conduction electrons also exhibit
an antiferromagnetic order through the Kondo couplings with the
localized spins. One can see this antiferromagnetic order from
\bqa &&
\langle{c}_{i\sigma}^{\dagger}\tau_{3\sigma\sigma'}c_{i\sigma'}
\rangle =
\langle{\psi}_{i\alpha}^{\dagger}U_{i\alpha\sigma}^{\dagger}\tau_{3\sigma\sigma'}U_{i\sigma'\beta}\psi_{i\beta}
\rangle \approx
\langle{\psi}_{i\alpha}^{\dagger}\psi_{i\beta}\rangle
\langle{U}_{i\alpha\sigma}^{\dagger}\tau_{3\sigma\sigma'}U_{i\sigma'\beta}
\rangle \nn && =
\langle{\psi}_{i\uparrow}^{\dagger}\psi_{i\uparrow} -
\psi_{i\downarrow}^{\dagger}\psi_{i\downarrow}\rangle \langle
z_{i\uparrow}^{\dagger}z_{i\uparrow} -
z_{i\downarrow}^{\dagger}z_{i\downarrow}\rangle \not= 0 ,  \eqa if
the easy axis anisotropy is assumed. In the easy plane limit one
finds
$\langle{c}_{i\sigma}^{\dagger}\tau_{1\sigma\sigma'}c_{i\sigma'}
\rangle \not= 0$ or
$\langle{c}_{i\sigma}^{\dagger}\tau_{2\sigma\sigma'}c_{i\sigma'}
\rangle \not= 0$. Since the conduction electrons form an insulator
due to the Fermi-nesting at half filling, the phase of the
conduction electrons is an antiferromagnetic insulator.

When the localized spins are in a disordered phase, the
antiferromagnetic order of the conduction electrons also vanishes
as $\langle{z}_{i\sigma}\rangle = 0$. As a result, for $J_{K}
> J_{Kc}$ the phase of the conduction electrons is identified
as a Kondo insulator because the conduction electrons are still
gapped due to the Fermi-nesting. In the Kondo insulator the
origin of the excitation gap is the Kondo hybridization, not the
antiferromagnetic ordering. Figure 1 shows that mean field analysis for the
effective action of the Kondo lattice model exhibits  a second order
phase transition from an antiferromagnetic insulator to a Kondo
insulator, as increases the Kondo coupling strength $J_{K}$.
The possible mean field phases at half filling are summarized
in Table \ref{T1}.

\begin{table*}
\caption{Quantum phases at half filling in the Kondo lattice
model}
\begin{tabular}{cccccccc}
\hline $J_{K} < J_{Kc}$ & $J_{K} > J_{Kc}$ \nn Antiferromagnetic
insulator & Paramagnetic insulator \nn \hline
$\langle{z}_{i\sigma}\rangle \not= 0$ &
$\langle{z}_{i\sigma}\rangle = 0$ \nn
$\langle{c}_{i\alpha}^{\dagger}{\tau}_{3\alpha\beta}c_{i\beta}\rangle
\not= 0$ &
$\langle{c}_{i\alpha}^{\dagger}{\tau}_{3\alpha\beta}c_{i\beta}\rangle
= 0$  \nn
$\langle{\psi}_{i\alpha}^{\dagger}{\tau}_{3\alpha\beta}\psi_{i\beta}\rangle
\not= 0$ &
$\langle{\psi}_{i\alpha}^{\dagger}{\tau}_{3\alpha\beta}\psi_{i\beta}\rangle
\not= 0$ \nn \hline \label{T1}
\end{tabular}
\end{table*}

Although the continuous quantum transition between the two
insulating phases was obtained in the mean field approximation, it
should be considered as an artifact of the mean field analysis
because instanton excitations of compact U(1) gauge
fields\cite{Instanton} cause confinement of the massive spinons
and chargons in the Kondo insulating phase beyond the mean field
level. From the seminal work of Fradkin and Shenker\cite{Fradkin}
we know that there can be no phase transition between the Higgs
and confinement phases. The order parameter discriminating the
Higgs phase from the confinement one has not been known
yet.\cite{NaLee} In this respect only a crossover behavior is
expected. In this study the antiferromagnetic state
corresponds to the Higgs phase because the phase is characterized
by the spinon condensation, while the Kondo insulating state
coincides with the confinement phase. Applying Fradkin and
Shenker's result to the present problem, we conclude that the
second order phase transition turns into a crossover between the
antiferromagnetic insulator and the Kondo insulator. This
crossover picture is reasonable, considering that the chargon
excitations are gapped in both phases.

Note that the present spin decomposition is not allowed for all
values of $J_{K}/D$ because the renormalized hopping integrals
$tF_{r}$ and $tE_{r}$ for the spinons and chargons, respectively,
become zero above a certain value of $J_{K}/D$ (Fig. 1) in the Kondo
insulator. Solving the mean field equations (21) in the limit of
$E_{r} \rightarrow 0$ and $F_{r} \rightarrow 0$, one can determine
the value of $J_{K}S/D$ resulting in $E_{r} = 0$ and $F_{r} = 0$ from the
following conditions \bqa && \left(F_t
E_t+\frac{2}{\pi}\right)F_t=1 , ~~~ \left(F_t
E_t-\frac{2}{\pi}\right)E_t=\frac{J_K S}{J} , ~~~  18\frac{J_K S
J}{D^2}=\frac{F_t}{E_t} ,~~~ \lambda = \frac{J}{F_{t}} . \nonumber
\eqa These equations give $J_{K}S/D = 0.19$ and $E_{t} = 8.99$ for
$J/D = 0.01$, and $J_{K}S/D = 0.13$ and $E_{t} = 2.27$ for $J/D =
0.1$. Both the spinon and chargon bands become flat above this
Kondo coupling strength, causing these particles localized with
$\langle{z}_{i\sigma}\rangle = 0$. We believe that this
localization originates from our strong coupling approach. We
interpret the localization as the breakdown of our spin
decomposition.

\subsection{Heavy fermion metal}

Now we consider a hole-doped case where the Fermi-nesting is
destroyed. We can expect the metallic behavior of chargons.
Introducing the electron chemical potential, we obtain the
self-consistent mean field equations for the chargon sector \bqa
&& JF_{t} =
\sum'_{k}\Bigl[\frac{2E_{k}^{\psi}}{E_{t}^{2}}\Bigl(n_{f}(-\frac{E_{k}^{\psi}-\mu}{E_{t}})
- n_{f}(\frac{E_{k}^{\psi}+\mu}{E_{t}})\Bigr) +
\frac{2\mu}{E_{t}^{2}}\Bigl(2 -
n_{f}(-\frac{E_{k}^{\psi}-\mu}{E_{t}}) -
n_{f}(\frac{E_{k}^{\psi}+\mu}{E_{t}})\Bigr) \Bigr] \nn &&
+\frac{2J}{\pi{E}_{t}} \nn && =_{(T \rightarrow 0)}
\frac{1}{4D}\int_{-D}^{D}d{\epsilon}
\Bigl[\frac{2\sqrt{E_{r}^{2}\epsilon^{2}+(J_{K}S)^2}}{E_{t}^{2}}
\Bigl(\Theta(\frac{\sqrt{E_{r}^{2}\epsilon^{2}+(J_{K}S)^2}-\mu}{E_{t}})
-
\Theta(-\frac{\sqrt{E_{r}^{2}\epsilon^{2}+(J_{K}S)^2}+\mu}{E_{t}})
\Bigr) \nn &&
 + \frac{2\mu}{E_{t}^{2}}\Bigl(2 -
\Theta(\frac{\sqrt{E_{r}^{2}\epsilon^{2}+(J_{K}S)^2}-\mu}{E_{t}})
-
\Theta(-\frac{\sqrt{E_{r}^{2}\epsilon^{2}+(J_{K}S)^2}+\mu}{E_{t}})
\Bigr)\Bigr]+ \frac{2J}{\pi{E}_{t}}, \nn && DF_{r} =
\frac{2E_{r}}{E_{t}}\sum'_{k}\frac{\epsilon_{k}^{\psi{2}}}{E_{k}^{\psi}}\Bigl(n_{f}(-\frac{E_{k}^{\psi}-\mu}{E_{t}})
- n_{f}(\frac{E_{k}^{\psi}+\mu}{E_{t}})\Bigr) \nn && =_{(T
\rightarrow 0)} \frac{E_{r}}{2DE_{t}}\int_{-D}^{D}d\epsilon
\frac{\epsilon^{2}}{\sqrt{E_{r}^{2}\epsilon^{2}+(J_{K}S)^2}}\Bigl(\Theta(\frac{\sqrt{E_{r}^{2}\epsilon^{2}+(J_{K}S)^2}-\mu}{E_{t}})
-
\Theta(-\frac{\sqrt{E_{r}^{2}\epsilon^{2}+(J_{K}S)^2}+\mu}{E_{t}})
\Bigr) , \nn && 1 - \delta = 2\sum'_{k}\Bigl(2 -
n_{f}(-\frac{E_{k}^{\psi}-\mu}{E_{t}}) -
n_{f}(\frac{E_{k}^{\psi}+\mu}{E_{t}})\Bigr) \nn && =_{(T
\rightarrow 0)} \frac{1}{2D}\int_{-D}^{D}d\epsilon \Bigl(2 -
\Theta(\frac{\sqrt{E_{r}^{2}\epsilon^{2}+(J_{K}S)^2}-\mu}{E_{t}})
-
\Theta(-\frac{\sqrt{E_{r}^{2}\epsilon^{2}+(J_{K}S)^2}+\mu}{E_{t}})
\Bigr) . \eqa  The mean field equations in the spinon sector
remains the same as those in Eq. (20). One can recover Eq. (20)
for half filling by setting $\delta = 0$ and $\mu = 0$ in Eq.
(25).

In the doped case two kinds of phase transitions are expected to
appear. One occurs in the spinon sector, characterized by the
spinon condensation, thus associated with an antiferro- to para-
magnetic transition of the localized spins. The other can appear
in the chargon sector, not understood by condensation of an order
parameter since there is no order parameter in this fermion part.
The phase transition is an insulator to metal transition of the
chargon excitations, occurring when the gap in the chargon
spectrum vanishes. From the last equation in Eq. (25) one can see
how the chemical potential changes as a function of $\delta$ and
$J_{K}$, given by \bqa && \mu = - \sqrt{(E_{r}D)^{2}\delta^{2} +
(J_{K}S)^{2}} \eqa for $\delta > 0$ and $\mu = 0$ for $\delta =
0$. This means that as soon as holes are doped in the conduction
band, the chemical potential that lies between the upper and lower
"conduction" bands jumps to the lower band, thus metallic
properties of the conduction electrons appear.

The most important question in this study is how the antiferro- to
para- magnetic transition of the localized spins arises when the
conduction electrons become metallic away from half filling. This
quantum transition is driven by the spinon condensation.
Performing the momentum integrals in the first and second
equations in Eq. (25), we obtain the following expressions for the
mean field equations \bqa  JF_{t} &=& \frac{2J}{\pi{E}_{t}} +
\frac{\mu}{E_{t}^{2}}\Bigl(1 - \frac{\sqrt{\mu^{2} -
(J_{K}S)^{2}}}{DE_{r}}\Bigr) \nn &+&
\frac{1}{2DE_{t}^{2}E_{r}}\Big(DE_{r}\sqrt{D^2E_{r}^{2}+(J_KS)^2}+\mu
\sqrt{\mu^2-(J_KS)^2}
\nn &+&
(J_KS)^2\left[\sinh^{-1}(DE_{r}/J_KS)-\sinh^{-1}(\sqrt{\mu^2-(J_KS)^2}
/J_KS)\right]\Big),
\nn  DF_{r} &=&
\frac{1}{2DE_{t}E_{r}^{2}}\Big(DE_{r}\sqrt{D^2E_{r}^{2}+(J_KS)^2}+\mu
\sqrt{\mu^2-(J_KS)^2}
\nn &-&(J_KS)^2\left[\sinh^{-1}(DE_{r}/J_KS)-\sinh^{-1}(\sqrt{\mu^2-(J_KS)^2}
/J_KS)\right]\Big).  \eqa Using the equations for
$E_{t}$, $E_{r}$, $\lambda$ in Eq. (21) and Eq. (27) with Eq.
(26), we can find the quantum critical point associated with the
magnetic transition, given by \bqa &&
F_{tc}^{3}\Bigl(\sqrt{1+(\frac{3J_{Kc}S}{D})^{2}} -
\sqrt{\delta^{2}+(\frac{3J_{Kc}S}{D})^{2}}\Bigr) =
\frac{24}{\pi^2}\frac{J}{D}F_{tc}^{2} -
\frac{28}{\pi}\frac{J}{D}F_{tc}+8\frac{J}{D} , \nn &&
F_{tc}^{3}\Bigl[\Bigl(\frac{3J_{Kc}S}{D}\Bigr)^{2}\Bigl(\sinh^{-1}\Bigl(\frac{1}{3J_{Kc}S/D}\Bigr)
- \sinh^{-1}\Bigl(\frac{\delta}{3J_{Kc}S/D}\Bigr)\Bigr) -
(1-\delta)\sqrt{\delta^{2}+(\frac{3J_{Kc}S}{D})^{2}} \Bigr]
\nn && =
\frac{24}{\pi^2}\frac{J}{D}F_{tc}^{2} -
\frac{20}{\pi}\frac{J}{D}F_{tc} + \frac{8}{3}\frac{J}{D} .
\eqa Note that Eq. (28) is not reduced to Eq. (22) in the $\delta
\rightarrow 0$ limit because there is a chemical potential jump at
half filling.

\begin{figure}[t]
\vspace*{10cm} \includegraphics{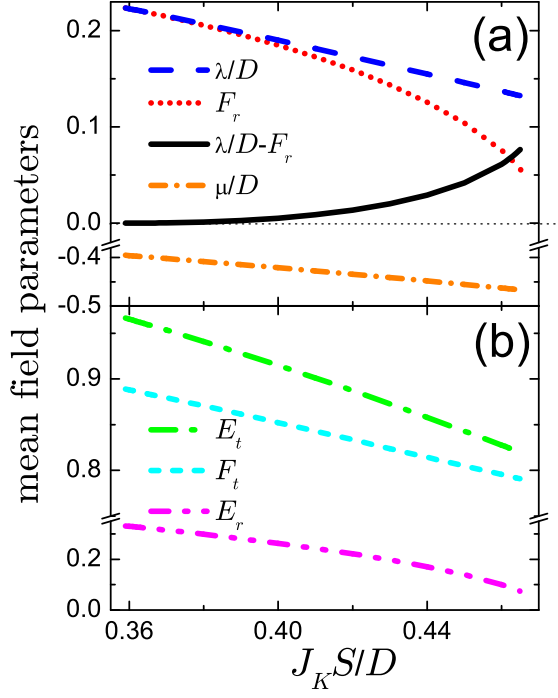} \vspace*{0cm} \caption{(Color
online) The mean field parameters away from half filling ($\delta
= 0.5$) are shown as varying $J_{K}S/D$ with $J/D = 0. 1$. The
quantum critical point defined by $\lambda_{c}/D - F_{rc} = 0$ in
(a) coincides exactly with the analytic result $J_{Kc}S/D \approx
0. 36$. The chemical potential agrees well with the analytic
result $\mu = - \sqrt{(E_{r}D)^{2}\delta^{2} + (J_{K}S)^{2}}$.
Even away from half filling the effective hopping parameters
$F_{r}$ and $E_{r}$ also vanish at a certain value of the Kondo
coupling beyond its critical point. }
\end{figure}

Let us consider the doped case with $\delta = 0.5$.
For $J/D = 0.1$ we find
$J_{Kc}S/D \approx 0.36$, indicating the existence of the phase
transition from an antiferromagnetic metal to a paramagnetic metal
away from half filling. Remember that there is no phase transition
at half filling when $J/D = 0.1$. We identify this
antiferromagnetic metal to paramagnetic metal transition as the
quantum phase transition in the Kondo lattice model. Possible mean
field phases away from half filling are summarized in Table
\ref{T2}, basically the same as Table \ref{T1} except that the
phases are metallic rather than insulating. The mean field
parameters away from half filling is shown in Fig. 2, where  $J/D=0.1$.
Note that the transition point
in Fig. 2, obtained by solving the mean field equations (21),
(26), and (27) numerically, agrees completely with the analytic
calculation.

A standard way of interpreting the quantum transition uses the ground
state wave function. For $J_{K} < J_{Kc}$, since $\langle{z}_{i\sigma}\rangle \not=
0$ and $\langle{\psi}_{i\alpha}^{\dagger}{\tau}_{3\alpha\beta}\psi_{i\beta}\rangle
\not= 0$, the ground state
$|\vec{n}_{i}\rangle$ for a localized spin at site $i$ can be written
as $|\vec{n}_{i}\rangle = n_{AF}|AF\rangle + n_{KS}|KS\rangle$,
where $|AF\rangle$ is the antiferromagnetic state with weight
$n_{AF}$, and $|KS\rangle$ the Kondo singlet state with weight
$n_{KS}$. We emphasize again that the Kondo hybridization still
exists for $J_{K} < J_{Kc}$. The ground state of the conduction
electrons for $J_{K} < J_{Kc}$ is given by $|c_{i\sigma}\rangle =
c_{AF}|AF\rangle + c_{KS}|KS\rangle$ in the same way, where the
antiferromagnetism and Kondo hybridization result from the Kondo
interaction. Increasing the Kondo coupling strength, the
antiferromagnetic long range order vanishes due to the Kondo
hybridization, thus the ground state for the localized spin at site
$i$ turns into $|\vec{n}_{i}\rangle = |KS\rangle$. In the same way
the ground state for the conduction electrons is given by
$|c_{i\sigma}\rangle = |KS\rangle$.

This discussion implies that fluctuations of the Kondo singlets
are not critical in this transition, and only antiferromagnetic
spin fluctuations are critical to drive the quantum transition via
the Kondo interaction. This picture is consistent with the single
impurity problem, where the Kondo coupling always causes the Kondo
singlet ground state. The presence of the Kondo singlets in both
phases originates from the strong coupling approach, where the
Kondo coupling term is solved first, thus allowing the Kondo
singlets in both phases. This leads us to conclude that the volume
change of Fermi surface should be continuous, which can be checked
from the fact that the chemical potential varies continuously
across the transition.

\begin{table*}
\caption{Quantum phases away from half filling in the Kondo
lattice model}
\begin{tabular}{cccccccc}
\hline $J_{K} < J_{Kc}$ & $J_{K} > J_{Kc}$ \nn Antiferromagnetic
metal & Heavy fermion metal \nn \hline
$\langle{z}_{i\sigma}\rangle \not= 0$ &
$\langle{z}_{i\sigma}\rangle = 0$ \nn
$\langle{c}_{i\alpha}^{\dagger}{\tau}_{3\alpha\beta}c_{i\beta}\rangle
\not= 0$ &
$\langle{c}_{i\alpha}^{\dagger}{\tau}_{3\alpha\beta}c_{i\beta}\rangle
= 0$  \nn
$\langle{\psi}_{i\alpha}^{\dagger}{\tau}_{3\alpha\beta}\psi_{i\beta}\rangle
\not= 0$ &
$\langle{\psi}_{i\alpha}^{\dagger}{\tau}_{3\alpha\beta}\psi_{i\beta}\rangle
\not= 0$ \nn \hline \label{T2}
\end{tabular}
\end{table*}

However, there can exist another solution for the mean field
equations (21), (26), and (27). If we assume $E_{r} = 0$ and
$F_{r} = 0$ for $J_KS/D$ larger than  the quantum critical point $J_{Kc}S/D$, these
mean field equations become \bqa && \left(F_{t}E_{t}+
\frac{2}{\pi}\right){F_t}=1 , ~~~~~ F_{t}E_{t} = \frac{2}{\pi} ,
~~~~~ \lambda = \frac{J}{F_t} , \nonumber \eqa yielding $F_{t} =
\pi/4$, $E_{t} = 8/\pi^2$, and $\lambda = 4J/\pi$. This solution
has an interesting physical interpretation although our
spin-decomposed effective action for the Kondo lattice model is
not available above the value of $J_K$ where $E_{r} \rightarrow 0$ and
$F_{r} \rightarrow 0$. Note
that this solution cannot be compatible with the spinon
condensation because the spinon condensation occurs when
$(D/J)F_{t}F_{r} \geq 2$ is satisfied. Thus, it can be
allowed only in the paramagnetic phase. Approaching the quantum
critical point from the antiferromagnetic phase, the effective
hopping parameters remain finite, i.e., $E_{rc} \not= 0$ and
$F_{rc} \not= 0$ while they are zero approaching the quantum
critical point from the paramagnetic phase. There is discontinuity
for $E_{r}$ and $F_{r}$, given by $\Delta{E}_{r} = E_{r}(J_{K}
\rightarrow J_{Kc}-0) - E_{r}(J_{K}
\rightarrow J_{Kc}+0) = E_{rc}$ and
$\Delta{F}_{r} = F_{r}(J_{K}
\rightarrow J_{Kc}-0) - F_{r}(J_{K}
\rightarrow J_{Kc}+0) = F_{rc}$. We expect that this abrupt change in
the hopping parameters may be related with discontinuity in the
volume change of the Fermi surface\cite{Review_theory} even if the
quantum transition is the second order described by the spinon
condensation. Although the delocalized solution ($E_{r} \not= 0$,
$F_{r} \not= 0$) near the quantum critical point is expected to
be a genuine solution in the mean field level, it will be
meaningful to consider this localization solution, considering
gauge fluctuations beyond the mean field approximation. Later, we will
comment on this issue shortly. Since we don't have any clear
physical picture for this localization, we do not go further based
on this mean field solution.

\section{Beyond the mean field approximation}

\subsection{Effective field theory}

To examine non-Fermi liquid physics near the quantum critical
point, it is necessary to obtain an effective continuum action of
Eq. (15). It is important to introduce non-perturbative effects of
Kondo interactions in the continuum action. We rewrite the mean
field action for the fermion sector in Eq. (16), and diagonalize
it as \bqa && S^{MF}_{\psi} =
\sum_{\omega_{n}}\sum_{k\sigma}\left(
\begin{array}{cc} \psi_{\omega_{n}k\sigma}^{\dagger} &
\psi_{\omega_{n}k+Q\sigma}^{\dagger} \end{array} \right) \left(
\begin{array}{cc} iE_{t}\omega_{n} - \mu + E_{r}\epsilon_{k}^{\psi} & \sigma{J}_{K}S \\
\sigma{J}_{K}S & iE_{t}\omega_{n} - \mu +
E_{r}\epsilon_{k+Q}^{\psi} \end{array} \right)\left(
\begin{array}{c} \psi_{\omega_{n}k\sigma} \\
\psi_{\omega_{n}k+Q\sigma} \end{array} \right) \nn && =
\sum_{\omega_{n}}\sum_{k\sigma}\left(
\begin{array}{cc} \eta_{+\omega_{n}k\sigma}^{\dagger} &
\eta_{-\omega_{n}k\sigma}^{\dagger} \end{array} \right) \left(
\begin{array}{cc} iE_{t}\omega_{n} - \mu + E_{k}^{\eta} & 0 \\
0 & iE_{t}\omega_{n} - \mu - E_{k}^{\eta}
\end{array} \right)\left(
\begin{array}{c} \eta_{+\omega_{n}k\sigma} \\
\eta_{-\omega_{n}k\sigma} \end{array} \right) , \eqa where the
$\eta_{\pm\omega_{n}k\sigma}$ fermions are given by the unitary
transformation of the $\psi_{\omega_{n}k\sigma}$ fermions in the
following way \bqa && \left(
\begin{array}{c} \eta_{+\omega_{n}k\sigma} \\
\eta_{-\omega_{n}k\sigma} \end{array} \right) =  \left(
\begin{array}{cc} \cos\vartheta_{\omega_{n}k} & -
\sigma\sin\vartheta_{\omega_{n}k} \\
\sigma\sin\vartheta_{\omega_{n}k} & \cos\vartheta_{\omega_{n}k}
\end{array} \right)\left(
\begin{array}{c} \psi_{\omega_{n}k\sigma} \\
\psi_{\omega_{n}k+Q\sigma} \end{array} \right) .  \eqa Here
$E_{k}^{\eta} = \sqrt{ (E_{r} \epsilon_{k}^{\psi} )^2 + (J_K S)^2
}$ is the quasiparticle energy obtained before, and
$\cos\vartheta_{\omega_{n}k}$, $\sin\vartheta_{\omega_{n}k}$ are
coherence factors, given by $\cos^{2}\vartheta_{\omega_{n}k} =
\frac{1}{2}\Bigl[1 +
\frac{E_{r}\epsilon_{k}^{\psi}}{E_{k}^{\eta}}\Bigr]$ and
$\sin^{2}\vartheta_{\omega_{n}k} = \frac{1}{2}\Bigl[1 -
\frac{E_{r}\epsilon_{k}^{\psi}}{E_{k}^{\eta}}\Bigr]$.

Expanding the quasiparticle band in the long wave length limit, we
obtain \bqa && E_{k}^{\eta} = J_{K}S\sqrt{1 + \Bigl(\frac{E_{r}
\epsilon_{k}^{\psi}}{J_{K}S} \Bigr)^2}
\approx J_{K}S\Bigl\{1 +
\frac{1}{2} \Bigl(\frac{E_{r} \epsilon_{k}^{\psi}}{J_{K}S}
\Bigr)^2\Bigr\}
\nn &&~~~
 \approx J_{K}S + \frac{2(E_{r}t)^{2}}{J_{K}S} -
\frac{2(E_{r}t)^{2}}{J_{K}S}(k_{x}^{2} + k_{y}^{2}) + {\cal
O}(k^4) , \eqa where the terms beyond the fourth order are ignored
in the long wave length limit. Inserting the above into Eq. (29)
and performing the Fourier transformation, one can obtain a low
energy continuum action for the $\eta_{-\sigma}$ fermions \bqa &&
S_{\psi} = \int_{0}^{\beta}{d\tau}\int{d^2r}\Bigl[ \sum_{\sigma}
\Bigl( \eta^{\dagger}_{-\sigma}(E_{\eta}[\partial_{\tau} -
i\sigma{a}_{\tau}] - \mu - iA_{\tau})\eta_{-\sigma}
\nn &&~~~~
+
\frac{1}{2M_{\eta}}|(\vec{\nabla} - i\sigma\vec{a} -
i\vec{A})\eta_{-\sigma}|^{2} \Bigr) \Bigr], \eqa
where $M_{\eta}^{-1} \equiv (DE_{r})^{2}/8J_{K}S$ and
$E_{t}$ is replaced by $E_{\eta}$.
Note that the effective mass $M_{\eta}$ of the renormalized conduction
electrons $\eta_{-\sigma}$ is proportional to the Kondo coupling.
The U(1) gauge field $a_{\mu}$ is introduced by shifting the three
momentum $k_{\mu}$ as $k_{\mu} - \sigma a_{\mu}$. The empty high
energy band for the $\eta_{+\sigma}$ fermions is ignored in the
low energy limit.

Performing the continuum approximation for the boson sector in Eq.
(15), the resulting effective field theory is given by   \bqa &&
{S}_{eff} = S_{\eta} + S_{z} , \nn && S_{\eta} =
\int_{0}^{\beta}{d\tau}\int{d^2r}\Bigl[ \sum_{\sigma} \Bigl(
\eta^{\dagger}_{\sigma}(E_{\eta}[\partial_{\tau} -
i\sigma{a}_{\tau}] - \mu - iA_{\tau})\eta_{\sigma} +
\frac{1}{2M_{\eta}}|(\vec{\nabla} - i\sigma\vec{a} -
i\vec{A})\eta_{\sigma}|^{2} \Bigr) \Bigr] , \nn && S_{z} =
\int_{0}^{\beta}{d\tau}\int{d^2r}\Bigl[
\sum_{\sigma}\Bigl(\frac{F_{z}}{J}|(\partial_{\tau} -
ia_{\tau})z_{\sigma}|^{2} + \frac{1}{2M_{z}}|(\vec{\nabla} -
i\vec{a})z_{\sigma}|^{2} + m_{z}^{2}|z_{\sigma}|^{2}\Bigr) \nn &&
+ \frac{u_{z}}{2}(\sum_{\sigma}|z_{\sigma}|^{2})^{2} +
w_{z}|z_{\uparrow}|^{2}|z_{\downarrow}|^{2} \Bigr]  , \eqa where
$M_{z}^{-1} \equiv tF_{r}/2$ and $F_{t}$ is replaced by $F_{z}$ in
the spinon part. Here the "$-$" symbol in $\eta_{-\sigma}$ field
is omitted for a simple notation. The unimodular constraint in the
spinon sector is softened via their local interactions $u_{z}$.
The $w_{z}$ term is phenomenologically introduced, associated with
the easy-plane anisotropy.

The above effective action can be simplified via the following
scale transformation \bqa && \tau' = \frac{\tau}{\sqrt{F_{z}/J}} ,
~~~~~~~ {\vec r}' = \sqrt{2M_{z}} {\vec r} . \eqa Performing the
field-transformation for the boson sector accordingly, \bqa &&
a_{\tau}' = \sqrt{F_{z}/J}a_{\tau} , ~~~~~~~ {\vec a}_{}' =
\frac{\vec{a}_{}}{\sqrt{2M_{z}}} , ~~~~~~~ z_{\sigma}'
=\frac{(F_{z}/J)^{1/4}}{\sqrt{2M_{z}}}z_{\sigma} , \eqa the
effective spinon action is given by \bqa && S_{z} =
\int_{0}^{\beta'}{d\tau'}\int{d^2r'}\Bigl[
\sum_{\sigma}\Bigl(|(\partial_{\tau'} -
ia_{\tau}')z_{\sigma}'|^{2} + |(\vec{\nabla}' -
i\vec{a}')z_{\sigma}'|^{2} + m_{z}^{'2}|z_{\sigma}'|^{2}\Bigr) \nn
&& + \frac{u_{z}'}{2}(\sum_{\sigma}|z_{\sigma}'|^{2})^{2} +
w_{z}'|z_{\uparrow}'|^{2}|z_{\downarrow}'|^{2} \Bigr] \eqa with
\bqa \beta' = \frac{\beta}{\sqrt{F_{z}/J}},~~~ m_{z}^{'2} =
m_{z}^{2} , ~~~ u_{z}' = \frac{2M_{z}}{(F_{z}/J)^{1/2}}u_{z} , ~~~
w_{z}' = \frac{2M_{z}}{(F_{z}/J)^{1/2}}w_{z} . \nonumber \eqa The
effective chargon action can also be obtained as \bqa && S_{\eta}
= \int_{0}^{\beta'}{d\tau'}\int{d^2r'}\Bigl[ \sum_{\sigma} \Bigl(
\eta^{'\dagger}_{\sigma}(E_{\eta}'[\partial_{\tau'} -
i\sigma{a}_{\tau}'] - \mu' - iA_{\tau}')\eta_{\sigma}' +
\frac{1}{2M_{\eta}'}|(\vec{\nabla}' - i\sigma\vec{a}' -
i\vec{A}')\eta_{\sigma}'|^{2} \Bigr) \Bigr] \nn \eqa with the
scale transformation for the fermion sector \bqa && E_{\eta}' =
\frac{E_{\eta}}{\sqrt{F_{z}/J}} , ~~~~~~~ M_{\eta}' =
\frac{M_{\eta}}{2M_{z}} , ~~~~~~~ {\vec A}_{}' =
\frac{\vec{A}_{}}{\sqrt{2M_{z}}} , ~~~~~~~
\eta_{\sigma}' =\frac{(F_{z}/J)^{1/4}}{\sqrt{2M_{z}}}\eta_{\sigma},~~~\nonumber\\
&&\mu' = \mu,~~~ A_{\tau}' = A_{\tau}.
\eqa

As a result, we find the effective field theory \bqa && {\cal
S}_{eff} = \int_{0}^{\beta}{d\tau}\int{d^2r}\Bigl[ \sum_{\sigma}
\Bigl( \eta^{\dagger}_{\sigma}(E_{\eta}[\partial_{\tau} -
i\sigma{a}_{\tau}] - \mu - iA_{\tau})\eta_{\sigma} +
\frac{1}{2M_{\eta}}|(\vec{\nabla} - i\sigma\vec{a} -
i\vec{A})\eta_{\sigma}|^{2} \Bigr) \Bigr] \nn && +
\int_{0}^{\beta}{d\tau}\int{d^2r}\Bigl[
\sum_{\sigma}\Bigl(|(\partial_{\tau} - ia_{\tau})z_{\sigma}|^{2} +
|(\vec{\nabla} - i\vec{a})z_{\sigma}|^{2} +
m_{z}^{2}|z_{\sigma}|^{2}\Bigr) +
\frac{u_{z}}{2}(\sum_{\sigma}|z_{\sigma}|^{2})^{2} +
w_{z}|z_{\uparrow}|^{2}|z_{\downarrow}|^{2} \Bigr] \nn && +
\frac{1}{2}\sum_{q,\omega_{n}}\Bigl(\frac{q^2}{g} +
\frac{E_{\eta}}{v_{\eta}}\frac{|\omega_{n}|}{q}\Bigr)\Bigl(\delta_{ij}
- \frac{q_{i}q_{j}}{q^2}\Bigr)a_{i}a_{j} , \eqa where the prime
symbol is omitted for a simple notation. In the gauge action the
former with an internal gauge charge $g$ of the $\eta_{\sigma}$
and $z_{\sigma}$ particles is the Maxwell term resulting from high
energy fluctuations of the $\eta_{\sigma}$ and $z_{\sigma}$
particles.\cite{Gauge_dynamics,Kim_U1SL} The latter with the
quasiparticle "renormalization" $E_{\eta}$ and the Fermi velocity
of the chargons $v_{\eta}$ is the Landau damping term representing
dissipative dynamics of gauge fluctuations, which come from
particle-hole excitations of the $\eta_{\sigma}$ fermions near the
Fermi surface.\cite{Gauge_dynamics} Since the time component of
the gauge field mediates local interactions due to the
$\eta_{\sigma}$ polarization\cite{Gauge_dynamics} which are
irrelevant in the renormalization group sense, it can be ignored
in low energy limit.

\subsection{Antiferromagnetic metal}

For weak Kondo couplings $J_{K} < J_{Kc}$ ($m_{z}^{2} < 0$) the
bosonic spinons become condensed, leading to an antiferromagnetic
order of the localized spins. An antiferromagnetic metal appears
for the conduction electrons away from half filling. Gauge
fluctuations are massive due to the Anderson-Higgs mechanism, thus
safely ignored in the low energy limit. Although the gauge
fluctuations are irrelevant in the renormalization group sense,
they play an important role in confining the fermionic chargons
with the bosonic spinons to make usual conduction electrons. This
can be seen from the unitary gauge. If the easy plane limit $w_{z}
< 0$ is considered in Eq. (39), the spinons can be treated as
$z_{\sigma} = (1/\sqrt{2})e^{i\phi_{\sigma}}$. The unitary gauge
means $\tilde{a}_{\mu} = a_{\mu} - \partial_{\mu}\phi_{\uparrow}$,
causing an excitation gap for the $\tilde{a}_{\mu}$ fields. In
this unitary gauge the phase degrees of freedom appear in the
chargon sector, and these phase fields can be gauged away from the
gauge transformation $c_{\sigma} =
e^{-i\sigma\phi_{\uparrow}}\eta_{\sigma}$. Thus, low energy
excitations are antiferromagnons
$e^{i(\phi_{\uparrow}-\phi_{\downarrow})}$ in the localized spins
and electron excitations $c_{\sigma}$ in the renormalized
conduction band instead of the fermionic chargons $\eta_{\sigma}$.
These conduction electrons feel week staggered magnetic fields due
to the antiferromagnetic ordering of the localized spins.

\subsection{Quantum critical point}

As increases the Kondo coupling, the antiferromagnetic metal
approaches the quantum critical point where the antiferromagnetic
order vanishes. Critical boson fluctuations renormalize gauge
dynamics\cite{Kleinert} in the critical field theory \bqa && {\cal
S}_{c} = \int_{0}^{\beta}{d\tau}\int{d^2r}\Bigl[ \sum_{\sigma}
\Bigl( \eta^{\dagger}_{\sigma}(E_{\eta}\partial_{\tau} - \mu -
iA_{\tau})\eta_{\sigma} + \frac{1}{2M_{\eta}}|(\vec{\nabla} -
i\sigma\vec{a} - i\vec{A})\eta_{\sigma}|^{2} \Bigr) \Bigr] \nn &&
+ \frac{1}{2}\sum_{q,\omega_{n}}
\Bigl(\frac{E_{\eta}}{v_{\eta}}\frac{|\omega_{n}|}{q} +
\frac{N_z}{8}q\Bigr)\Bigl(\delta_{ij} -
\frac{q_{i}q_{j}}{q^2}\Bigr)a_{i}a_{j} , \eqa where $N_{z}=2$ is
the flavor number of the spinons. Integration for the critical
spinons should be understood in the renormalization group sense.
Since critical boson fluctuations yield a term linearly
proportional to momentum $q$ in the gauge action, the dynamical
critical exponent is obtained as $z = 2$.\cite{Dynamical_exponent}

In the random phase approximation\cite{Tsvelik,Rosch} the free
energy is given by \bqa && \frac{F}{V} =
\int\frac{d^2q}{(2\pi)^{2}}
\int\frac{d\omega}{2\pi}\coth\Bigl[\frac{\omega}{2T}\Bigr]
\tan^{-1}\Bigl[\frac{{\rm Im}D(q,\omega)}{{\rm
Re}D(q,\omega)}\Bigr] \nn && ~~~ \approx
\frac{1}{4\pi^{2}}\int_{T}^{\omega_c}{d\omega}\int_{0}^{\infty}{dq}q
\tan^{-1}\Bigl[\frac{8E_{\eta}}{v_{\eta}N_{z}}\frac{\omega}{q^2}\Bigr]
\nn && ~~~ = \frac{1}{4\pi^{2}}\frac{E_{\eta}}{v_{\eta}N_{z}}
\Bigl[
(3-2\ln\frac{8E_{\eta}}{v_{\eta}N_{z}}-2\ln\omega_{c})\omega_{c}^{2}
- (3-2\ln\frac{8E_{\eta}}{v_{\eta}N_{z}}-2\ln T)T^{2} \Bigr] ,
\eqa where $D(q,\omega) = \Bigl( -
i\frac{E_{\eta}}{v_{\eta}}\frac{\omega}{q} + \frac{N_z}{8}q\Bigr)$
is the gauge kernel in the real frequency $\omega$, and
$\omega_{c}$ is an energy cutoff. The specific heat is obtained to
be \bqa && C_{V} = - T \Bigl(\frac{\partial^{2}F}{\partial
T^{2}}\Bigr)_{V} = -
\frac{1}{\pi^{2}}\frac{E_{\eta}}{v_{\eta}N_{z}}\ln\frac{8E_{\eta}}{v_{\eta}N_{z}}T
-\frac{1}{\pi^{2}}\frac{E_{\eta}}{v_{\eta}N_{z}}T\ln T = -
\frac{1}{8\pi^{2}}\frac{T}{T_{0}}\ln\frac{T}{T_{0}} \eqa with an
energy scale $T_{0} =
\Bigl(\frac{8E_{\eta}}{v_{\eta}N_{z}}\Bigr)^{-1}$. Thus, the
specific heat coefficient $\gamma$ has a singular dependence
$\gamma = {C}_{V}/T \propto - \ln T$ in the $T \rightarrow 0$
limit. The logarithmic divergence also can be seen in the two
dimensional itinerant antiferromagnet, where its critical field
theory is characterized by the dynamical exponent $z =
2$.\cite{Rosch}

In the one-loop level the imaginary part of the fermion
self-energy is given by \bqa &&
\Sigma_{\eta}{''}(k,\epsilon_k^{\eta}) =
\int_{0}^{\infty}{d\omega}\int\frac{d^{D}k'}{(2\pi)^{D}}[n(\omega)
+ 1][1 - f(\epsilon_{k'}^{\eta})] \nn &&\times
\frac{(k+k')_{\alpha}(k+k')_{\beta}}{(2M_{\eta})^{2}}
\Bigl(\delta_{\alpha\beta} - \frac{q_{\alpha}q_{\beta}}{q^2}\Bigr)
{\rm Im}D(q,\omega)\delta(\epsilon_{k}^{\eta} -
\epsilon_{k'}^{\eta} - \omega) \nn && =
\frac{N_{\eta}}{2\pi{M}_{\eta}^2}\int_{0}^{\infty}{d\omega}
\int{d\epsilon'}{d\theta}\delta(\epsilon_k^{\eta}-\epsilon'
-\omega) [n(\omega) + 1][1 - f(\epsilon')]|{\bf k}
\times{\bf\hat{q}}|^2 \frac{v_{\eta}q\omega/E_{\eta}}{\omega^2 +
\Bigl(\frac{N_{z}v_{\eta}}{8E_{\eta}}\Bigr)^{2}q^4} \nn &&
=\frac{k_FN_\eta}{2\pi{M_{\eta}^{2}}}\int_{0}^{\epsilon^{\eta}_{k}}d\omega\int_{0}^{\infty}dq
\frac{v_{\eta}q\omega/E_{\eta}}{\omega^2 +
\Bigl(\frac{N_{z}v_{\eta}}{8E_{\eta}}\Bigr)^{2}q^4} =
\frac{k_{F}}{M_{\eta}^{2}}\frac{N_{\eta}}{N_{z}}
\epsilon_{k}^{\eta} , \eqa where $\epsilon_{k}^{\eta} =
k^2/2M_{\eta}$ is the energy dispersion of the $\eta_{\sigma}$
fermions, and $N_{\eta}$ is the density of states at the Fermi
energy. The scattering rate can be obtained from the  self-energy
expression with an additional $q^2$ factor in the
integrand.\cite{Gauge_dynamics} Since the imaginary part of the
fermion self-energy is linearly proportional to the fermion
dispersion, the dc conductivity\cite{Ioffe-Larkin} is given by
$\sigma_{c} \sim T^{-2}$.

\subsection{Region of strong Kondo couplings}

Increasing $J_{K}$ further from the quantum critical point, an
anomalous metallic phase appears. Integrating out the gapped
$z_{\sigma}$ excitations in Eq. (39), we obtain the Maxwell term
for gauge fluctuations \bqa && S_{NFL} =
\int_{0}^{\beta}{d\tau}\int{d^2r}\Bigl[ \sum_{\sigma} \Bigl(
\eta^{\dagger}_{\sigma}(E_{\eta}\partial_{\tau} - \mu -
iA_{\tau})\eta_{\sigma} + \frac{1}{2M_{\eta}}|(\vec{\nabla} -
i\sigma\vec{a} - i\vec{A})\eta_{\sigma}|^{2} \Bigr) \Bigr] \nn
&&~~~~~~~ + \frac{1}{2}\sum_{q,\omega_{n}} \Bigl(\frac{q^2}{g} +
\frac{E_{\eta}}{v_{\eta}}\frac{|\omega_{n}|}{q}\Bigr)\Bigl(\delta_{ij}
- \frac{q_{i}q_{j}}{q^2}\Bigr)a_{i}a_{j} , \eqa where the
dynamical critical exponent is $z = 3$. The effective field theory
of Eq. (44) is well known to cause non-Fermi liquid physics due to
scattering with massless gauge fluctuations. The imaginary part of
the fermion self-energy is given by $\omega^{2/3}$ at the Fermi
surface, implying that its real part also has the same frequency
dependence via the Kramer's Kronig relation, thus giving rise to a
non-Fermi liquid behavior.\cite{Chubukov_NFL} The coefficient
$\gamma$ of the specific heat is proportional to $- \ln T$ in
three spatial dimensions and $T^{-1/3}$ in two
dimensions.\cite{Senthil_Kondo} The dc conductivity is
proportional to $T^{-5/3}$ in three dimensions and $T^{-4/3}$ in
two dimensions.\cite{Gauge_dynamics} Note that the dynamical
exponent $z$ changes from $z = 2$ at the quantum critical point to
$z = 3$ in the non-Fermi liquid phase. Physical responses in the
Fermi liquid to non-Fermi liquid transition for two spatial
dimensions are summarized in Table \ref{T3}.

\begin{table*}
\caption{Physical response in the Fermi liquid to non-Fermi liquid
transition}
\begin{tabular}{cccccccc}
\hline & $J_{K} < J_{Kc}$ & $J_{K} \approx J_{Kc}$ & $J_{K} >
J_{Kc}$ \nn & Antiferromagnetic Fermi liquid & Quantum critical
point & Paramagnetic non-Fermi liquid \nn \hline $\gamma =
C_{v}/T$ & const. & $- \ln{T}$ & $T^{-1/3}$ \nn $\sigma_{dc}$ &
$T^{-2}$ & $T^{-2}$ & $T^{-4/3}$ \nn \hline \label{T3}
\end{tabular}
\end{table*}

\subsection{How to recover the Fermi liquid phase}

When the spinon excitations are gapped, they can be ignored in the
low energy limit.  Thus, if the gauge fluctuations are suppressed
in Eq. (44), Fermi liquid physics can be obtained. As the Kondo
coupling constant increases, the effective chargon mass $M_{\eta}
= 8J_{K}S/(DE_{r})^{2}$ becomes heavier and gauge fluctuations are
suppressed because $1/v_{\eta} \sim M_{\eta}$ in Eq. (44). This
may give rise to the Fermi liquid physics in the case of large
Kondo couplings. In this scenario the non-Fermi liquid is expected
to turn into the Fermi liquid continuously. In our mean field
analysis $M_{z} \rightarrow \infty$ and $M_{\eta} \rightarrow
\infty$ were found at the point where $E_r=0$ and $F_r=0$. Hence,
the spin decomposition scheme cannot cover the whole range of the
phase diagram so that we were not able to recover the Fermi liquid
phase.

There is another possibility associated with the
confinement-deconfinement transition due to the compactness of the
U(1) gauge field in the present problem. Note that we did not take
into account instanton excitations in the previous discussion. In
two space and one time dimensions there is no deconfined phase
owing to proliferation of instanton excitations when only gapped
fermion or boson excitations exist.\cite{Instanton} However, the
presence of gapless matter fields was recently argued to allow a
deconfined
phase.\cite{Nagaosa_Deconfinement,Senthil_DQCP,Kim_DQCP,Hermele_ASL,Kim_ASL,Kim_U1SL,Kim_AF}
Non-Fermi liquid phase corresponds to the deconfined phase which
gapless fermion ($\eta_{\sigma}$) excitations make stable against
instanton excitations.\cite{Kim_U1SL,Nagaosa_Deconfinement} The
present quantum critical point is identified as the deconfined
quantum critical point\cite{Senthil_DQCP} that can be stable due
to critical boson ($z_{\sigma}$) excitations\cite{Kim_DQCP} and
gapless fermion excitations.\cite{Kim_U1SL,Nagaosa_Deconfinement}
On the other hand, Fermi liquid corresponds to the confinement
phase. Instanton condensation leads to confinement between the
$\eta_{\sigma}$ fermion and the $z_{\sigma}$ boson to make an
electron $c_{\sigma} = U_{\sigma\sigma'}\eta_{\sigma'}$. We don't
know whether the non-Fermi liquid phase is stable against
instanton excitations or not. If it is stable, the
confinement-deconfinement transition corresponding to the
non-Fermi liquid to Fermi liquid transition would occur in the
strong Kondo coupling region. The nature of this transition may be
KT-like (Kosterlitz-Thouless).\cite{Kleinert_KT} On the contrary,
if the non-Fermi liquid phase is unstable against the confinement,
the parameter region of the non-Fermi liquid phase would shrink to
vanish. Then, the quantum critical point will coincide with the
point where localization occurs with $M_{z} \rightarrow \infty$
and $M_{\eta} \rightarrow \infty$.

\section{Discussion and Perspectives}

Our  approach has some analogies with that of Ref.
\onlinecite{Pepin_Kondo} where bosonic spinons are used  for the
localized spins, resulting in charged fermions for the Kondo
resonances. However, there are several important differences
between our approach and that of Ref. \onlinecite{Pepin_Kondo}.
Ref. \onlinecite{Pepin_Kondo} starts from the Kondo-Heisenberg
lattice model \bqa && H_{KHM} =
\sum_{k\sigma}\epsilon_{k}c_{k\sigma}^{\dagger}c_{k\sigma} +
J_{K}\sum_{i\sigma\sigma'}b_{i\sigma}^{\dagger}b_{i\sigma'}c_{i\sigma'}^{\dagger}c_{i\sigma}
+ J_{H}\sum_{\ij
\sigma\sigma'}b_{i\sigma}^{\dagger}b_{i\sigma'}b_{j\sigma'}^{\dagger}b_{j\sigma}
, \eqa where $c_{k\sigma}$ represents a conduction electron with
momentum $k$ and spin $\sigma$, and $\vec{S}_{i} =
\frac{1}{2}b_{i\sigma}^{\dagger}\vec{\tau}_{\sigma\sigma'}b_{i\sigma'}$
is the boson representation of  the localized spin $\vec{S}_{i}$.
Performing the HS transformation for the particle-hole channel in
the Kondo coupling term and the particle-particle channel in the
Heisenberg interaction term, Eq. (45) reads \bqa && H_{eff} =
\sum_{k\sigma}\epsilon_{k}c_{k\sigma}^{\dagger}c_{k\sigma} +
\sum_{i\sigma}(b_{i\sigma}^{\dagger}\chi_{i}^{\dagger}c_{i\sigma}
+ H.c. ) - \sum_{i}\frac{\chi_{i}^{\dagger}\chi_{i}}{J_{K}} \nn &&
~~~~~~ + \sum_{\ij \sigma}
(|\Delta_{ij}|e^{i\pi(i-j)}b_{i\sigma}^{\dagger}b_{j-\sigma}^{\dagger}
+ H.c.) - \sum_{\ij}\frac{|\Delta_{ij}|^{2}}{J_{H}} , \eqa where
the on-site bond variable $\chi_{i}$ is a Grassman field
associated with the Kondo resonance and the bond variable
$\Delta_{ij}$ is introduced to keep short range antiferromagnetic
correlations.

The crucial difference between two approaches lies in the HS
decoupling scheme of the Kondo interaction term; Ref.
\onlinecite{Pepin_Kondo} allows three kinds of matter fields that
correspond to two fermions $c_{k\sigma}$, $\chi_{i}$ and one boson
$b_{i\sigma}$ while our decomposition introduces only two kinds of
matter fields, one fermion $\psi_{i\sigma}$ and one boson
$z_{i\sigma}$. In fact, $\chi_{i}$ fermion corresponds to
$\psi_{i\uparrow}$ fermion while $\psi_{i\downarrow}$ fermion is
not allowed in Ref. \onlinecite{Pepin_Kondo}. Since $\chi_{i}$
field follows fermion statistics and the condensation of fermions
is not possible, the conventional mean field analysis in the
slave-boson approach\cite{Nagaosa_book} is not applicable.
Recently, there has been a progress in this spin-boson approach
for the single impurity problem based on the non-crossing
approximation scheme of the U(1) slave-boson
theory,\cite{Nagaosa_book} although its extension to the Kondo
lattice model has not been reported yet.\cite{Spinon_Kondo}

\begin{figure}[t]
\vspace*{7cm} \includegraphics{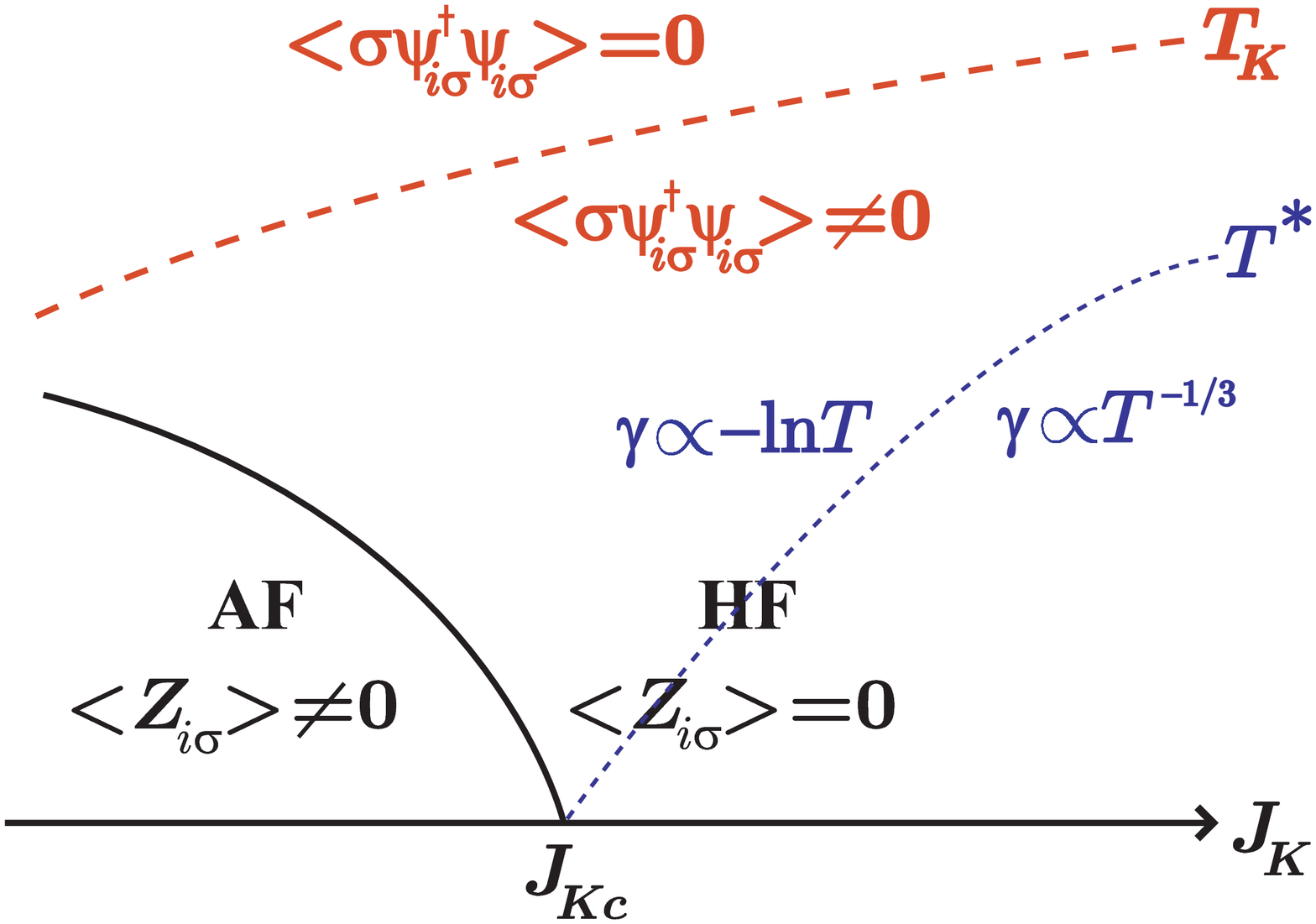} \vspace*{0cm} \caption{ (Color online) A schematic
phase diagram of the Kondo lattice model based on our effective
theory }
\end{figure}

A schematic phase diagram based on the effective Kondo action in
Eq. (15) is shown in Fig. 3, where horizontal axis is the Kondo
coupling strength and vertical is temperature. "AF" represents the
antiferromagnetic metal and "HF"  the heavy fermion metal. The
dashed line denoted by $T_{K}$ represents the Kondo temperature
for the Kondo singlets to form, where
$\langle\sigma\psi_{i\sigma}^{\dagger}\psi_{i\sigma}\rangle \not=
0$ below $T_{K}$ while
$\langle\sigma\psi_{i\sigma}^{\dagger}\psi_{i\sigma}\rangle = 0$
above $T_{K}$. The solid line shows the second order
antiferromagnetic transition related with the spinon condensation
in the present theory. This phase diagram is quite similar to that
of the HMM theory since the Kondo hybridization always exists in
both the antiferromagnetic and heavy fermion phases below the
Kondo temperature, and antiferromagnetic ordering is associated
with the quantum transition. However, an important difference
between these two theories can be found near the antiferromagnetic
quantum critical point.

In the HMM theory low energy elementary excitations at the quantum
critical point are critical antiferromagnetic fluctuations with
spin quantum number $1$, described by \bqa && S_{HMM} =
\frac{1}{2}\sum_{q,\omega_{n}}(\Gamma|\omega_{n}| +
q^{2})\vec{n}(q,\omega_{n})\cdot\vec{n}(-q,-\omega_n) +
V(|\vec{n}|) , \eqa where the damping term with a damping
coefficient $\Gamma$ comes from gapless electron excitations near
the Fermi surface and $V(|\vec{n}|) =
\int_{0}^{\beta}{d\tau}\int{d^2r} \frac{u_n}{2}|\vec{n}|^{4} +
...$ is an effective potential for the spin-fluctuation order
parameter $\vec{n}$.\cite{HMM} On the other hand, at the quantum
critical point of the present approach the critical antiferromagnetic fluctuations are
fractionalized into critical spinon excitations with spin quantum
number $1/2$ due to strong Kondo interactions. The critical
field theory is given by \bqa && S_{DQCP} =
\int_{0}^{\beta}{d\tau}\int{d^2r}\Bigl[
\sum_{\sigma}\Bigl(|(\partial_{\tau} - ia_{\tau})z_{\sigma}|^{2} +
|(\vec{\nabla} - i\vec{a})z_{\sigma}|^{2} \Bigr) \nn &&+
\frac{u_{z}}{2}(\sum_{\sigma}|z_{\sigma}|^{2})^{2} +
w_{z}|z_{\uparrow}|^{2}|z_{\downarrow}|^{2} \Bigr]  +
\frac{1}{2}\sum_{q,\omega_{n}}
\Bigl(\frac{E_{\eta}}{v_{\eta}}\frac{|\omega_{n}|}{q} +
\frac{N_z}{8}q\Bigr)\Bigl(\delta_{ij} -
\frac{q_{i}q_{j}}{q^2}\Bigr)a_{i}a_{j} , \eqa where the gapless
renormalized fermions $\eta_{\sigma}$ are integrated out, causing
dissipation in gauge dynamics. Here "DQCP" means the deconfined
quantum critical point\cite{Senthil_DQCP,Kim_DQCP} discussed in
the previous section.

It is interesting to see that both critical theories in Eqs. (47)
and (48) are characterized by the same dynamic critical exponent
$z = 2$. Thus, two critical theories show similar critical physics
although the low lying excitations are completely different.
Actually, the HMM theory can also explain the logarithmic
divergence in the specific heat coefficient in two spatial
dimensions.\cite{Review_theory} Since two spatial dimensions lie
in the upper critical dimension due to the dynamic critical
exponent, universal scaling for the spin susceptibility does not
appear in both critical theories. Thus, it is impossible to
compare an anomalous critical exponent in one theory with that of
the other. However, there exists one crucial difference; if we
have the deconfined quantum critical point, the non-Fermi liquid
metal as a deconfined critical phase can appear above the
deconfined quantum critical point ($J_{K} > J_{Kc}$) while this
non-Fermi liquid phase cannot be allowed in the HMM theory.
According to our effective field theory, it is clear that there
should be a crossover between the quantum critical region and
non-Fermi liquid phase as decreases temperature in $J_{K} >
J_{Kc}$. The crossover temperature $T^{*}$ depends on the spin gap
$\lambda - DF_{r}$ as $T^{*} \sim \lambda - DF_{r}$. This
crossover should appear in the upturn behavior of the specific
heat coefficient from $\gamma \sim - \ln T$ to $\gamma \sim
T^{-1/3}$ since the effective field theory changes from $z = 2$ to
$z = 3$ during the crossover. The upturn behavior was also
discussed in Ref. \onlinecite{Pepin_Kondo}, but the mechanism is
different.

Let us discuss dissipation in both the weak coupling critical
theory of Eq. (47) and the strong coupling one of Eq. (48). The
strong coupling theory may be derived from the weak coupling one
using the CP$^{1}$ representation. Remember that the O(3)
nonlinear $\sigma$ model can be mapped onto the U(1) gauge theory
within the CP$^{1}$ representation, as discussed before. The main
point of this CP$^{1}$ decomposition is how dissipative dynamics
of spin fluctuations ${\vec n}(r,\tau)$ in the HMM theory [Eq.
(47)] is transferred into that of gauge fluctuations
$\vec{a}(r,\tau)$ in the CP$^{1}$ gauge theory [Eq. (48)]. In the
context of the standard weak coupling theory order parameter
fluctuations directly couple to gapless fermion excitations near
the Fermi surface. As a result, dissipation effects in order
parameter fluctuations appear in the kinetic energy term. In our
strong coupling approach fractionalized order parameter
fluctuations do not couple to the gapless fermion excitations
directly. Instead, their couplings are realized indirectly via
gauge fluctuations. Dissipative dynamics of fractionalized order
parameter fluctuations is induced by damped gauge fluctuations
that result from the gapless fermion excitations near the Fermi
surface. The damped gauge fluctuations play an important role in
quantum critical physics. If fermion excitations are gapped and
the damping effects in gauge fluctuations are not taken into
account in Eq. (48), the quantum phase transition belongs to the
inverted XY (IXY) universality class in the case of large flavors
of fractionalized boson excitations.\cite{Kleinert} However, the
presence of dissipation in gauge excitations due to gapless
fermions changes the IXY universality class completely. Since the
dissipation results in the dynamic critical exponent  $z = 2$, the
spacial dimension $d =2$ becomes the upper critical dimension and
the nature of the quantum transition would be a mean field-like
type with logarithmic corrections.

Inserting the CP$^{1}$ representation $\vec{n} =
\frac{1}{2}z_{\sigma}^{\dagger}\vec{\tau}_{\sigma\sigma'}z_{\sigma'}$
into Eq. (47),\cite{Auerbach} Eq. (47) can be written in terms of
the bosonic spinons interacting with gauge fluctuations.
Unfortunately, the $|\omega_{n}|$ linear (damping) term prevents
obtaining a complete expression. Performing the HS transformation
for the damping term, dissipation in order parameter fluctuations
would appear in gauge fluctuations although the dissipative gauge
action is not given by that in Eq. (48). This implies that damping
effects due to gapless fermions are imposed in a different way for
the weak and strong coupling theories.

It is valuable to apply our spin decomposition to the one
dimensional Kondo lattice model in order to confirm that the
non-Fermi liquid metal with spin gap can be allowed. The effective
field theory in Eq. (39) will be applicable to the one dimensional
case. An important difference from the two dimensional case is
that dissipative dynamics in gauge fluctuations does not appear
because the gapless conduction fermions are described by massless
Dirac fermions near the Fermi points. In one dimensional effective
field theory strong quantum fluctuations coming from low
dimensionality do not allow the spinon condensation. Furthermore,
the spinon excitations are gapped because the Berry phase
contribution disappears due to the Kondo coupling. The most
crucial point in the one dimensional effective theory is that the
gapless Dirac fermions make gauge fluctuations massive,
\cite{Shankar} which can be seen  using the bosonization
technique. As a result, the gapped spinon excitations are
deconfined\cite{Kim_Kondo_1D} although the mechanism is different
from the two the dimensional case. Moreover, the gapless fermion
excitations exhibit strong superconducting correlations as the two
dimensional case. Our spin-gauge theory in Eq. (39) allows
superconducting instability because the spin-gauge fields mediate
attractive interactions between  $\eta_{\uparrow}$ and
$\eta_{\downarrow}$ fermions.  The $\sigma$ symbol in the gauge
coupling shows that the gauge charge of the $\eta_{\uparrow}$
fermion is opposite to that of the $\eta_{\downarrow}$ fermion. In
this respect the non-Fermi liquid phase with spin gap is the two
dimensional analogue of the one dimensional spin-gapped phase.

If SU(2) gauge fluctuations are taken into account in Eq. (11) instead of
U(1) gauge fluctuations, the non-Abelian nature of
SU(2) gauge fluctuations may not allow  the deconfined quantum
criticality and non-Fermi liquid phase.\cite{Instanton}
However,  there is no consensus for the confinement problem in the
context of the SU(2) gauge theory, as far as we know. If the
deconfined non-Fermi liquid phase turns into a confinement state,
spinons should be confined with chargons via SU(2) gauge fluctuations.
Instead, electron excitations are allowed
and the resulting phase may be the Fermi liquid.
The spinons and chargons are not meaningful
objects in the low energy limit.
However, the spinon excitations may emerge as broad spin spectrum (particle-hole
continuum) at high energies beyond multi-paramagnon scattering
according to the asymptotic freedom of the SU(2) gauge theory.\cite{Instanton}.

In this SU(2) gauge theoretic description the quantum critical
point would lie between the Higgs phase (antiferromagnetism) and
the confinement one (Fermi liquid), while it is between the Higgs
phase and the deconfinement one (non-Fermi liquid) in the present
U(1) gauge theory. It was argued that there is no phase transition
between the Higgs and confinement phases and they are smoothly
connected.\cite{Fradkin} Then, the spin-decomposition method in
the context of the SU(2) gauge theory cannot describe the quantum
phase transition of the Kondo lattice model. In this case another
order parameter should be considered to study the quantum phase
transition of the Kondo lattice model, for example, the
hybridization order parameter in the context of the slave-boson
theory.\cite{Senthil_Kondo,Kim_Kondo2}

In summary, we investigated the quantum phase transition from an
antiferromagnetic metal to a heavy fermion metal in the Kondo
lattice model. First, we diagonalized the Kondo coupling term in
the strong coupling approach.  Then, we derived the effective
Kondo action [Eq. (15)] and performed the mean field analysis [Eq.
(19)] to obtain the mean field phase diagram, showing the quantum
phase transition from the antiferromagnetic metal to the heavy
fermion metal. The Kondo term is always relevant so that the Kondo
hybridization persists even in the antiferromagnetic metal, which
means that fluctuations of the Kondo singlets are not critical in
the phase transition. The volume change of Fermi surface thus is
expected to be continuous across the transition. We found that
softening of antiferromagnetic spin fluctuations leads to the
quantum transition driven by the Kondo interaction in the strong
coupling approach. Beyond the mean field level we derived the
effective U(1) gauge theory [Eq. (39)] in terms of the
renormalized conduction electrons $\eta_{\sigma}$ and the
spin-fractionalized excitations $z_{\sigma}$ interacting via the
U(1) spin-gauge fields $a_{\mu}$. Our critical field theory
characterized by the critical exponent $z = 2$ can explain the
non-Fermi liquid physics such as $\gamma \sim - \ln T$ near the
quantum critical point. Furthermore, we showed that if our
scenario is applicable, there can exist a narrow region of the
non-Fermi liquid phase with spin gap near the quantum critical
point. We also discussed how the present theory can recover the
Fermi liquid phase, but this issue should be clarified near
future. Lastly, we commented on the superconducting instability
near the quantum critical point. This interesting possibility
remains as a future study.


\begin{thebibliography}{9}
\bibitem{Discrete_time} The time derivative term in the
discrete-time becomes in the limit of $\Delta\tau \rightarrow 0$
\bqa && - \sum_{\tau\tau'i}
\psi_{i\tau}^{\dagger\sigma}U_{i\tau}^{\dagger\sigma\alpha}
U_{i\tau'}^{\alpha\sigma'}\psi_{i\tau'}^{\sigma'} \approx -
\sum_{\tau{i}}
\psi_{i\tau}^{\dagger\sigma}U_{i\tau}^{\dagger\sigma\alpha}
(U_{i\tau}^{\alpha\sigma'} -
\frac{\Delta{U}_{i\tau}^{\alpha\sigma'}}{\Delta\tau}\Delta\tau)(\psi_{i\tau}^{\sigma'}
- \frac{\Delta\psi_{i\tau}^{\sigma'}}{\Delta\tau}\Delta\tau) \nn
&& = \sum_{\tau{i}} \Bigl(-
\psi_{i\tau}^{\dagger\sigma}\psi_{i\tau}^{\sigma} +
\psi_{i\tau}^{\dagger\sigma}U_{i\tau}^{\dagger\sigma\alpha}
\frac{\Delta{U}_{i\tau}^{\alpha\sigma'}}{\Delta\tau}\Delta\tau
\psi_{i\tau}^{\sigma'} + \psi_{i\tau}^{\dagger\sigma}
\frac{\Delta\psi_{i\tau}^{\sigma'}}{\Delta\tau}\Delta\tau -
\psi_{i\tau}^{\dagger\sigma}U_{i\tau}^{\dagger\sigma\alpha}
\frac{\Delta{U}_{i\tau}^{\alpha\sigma'}}{\Delta\tau}\Delta\tau
\frac{\Delta\psi_{i\tau}^{\sigma'}}{\Delta\tau}\Delta\tau \Bigr)
\nn && \approx -
J\sum_{\tau{i}}\Delta\tau\psi_{i\tau}^{\dagger\sigma}\psi_{i\tau}^{\sigma}
+
\sum_{\tau{i}}\Delta\tau\psi_{i\tau}^{\dagger\sigma}U_{i\tau}^{\dagger\sigma\alpha}
\frac{\Delta{U}_{i\tau}^{\alpha\sigma'}}{\Delta\tau}
\psi_{i\tau}^{\sigma'} +
\sum_{\tau{i}}\Delta\tau\psi_{i\tau}^{\dagger\sigma}
\frac{\Delta\psi_{i\tau}^{\sigma'}}{\Delta\tau} \nn && =
\int_{0}^{\beta}{d\tau}\sum_{i}\Bigl(\psi_{i\tau}^{\dagger\sigma}
(\partial_{\tau}\delta_{\sigma\sigma'} +
U_{i\tau}^{\dagger\sigma\alpha}
\partial_{\tau}{U}_{i\tau}^{\alpha\sigma'} )
\psi_{i\tau}^{\sigma'} -
J\psi_{i\tau}^{\dagger\sigma}\psi_{i\tau}^{\sigma} \Bigr) ,
\nonumber \eqa  where $\sum_{\tau}\Delta\tau$ is replaced with
$\int_{0}^{\beta}{d\tau}$ with $J = 1/\Delta\tau$, and the last
$J$ term can be absorbed in the chemical potential term.
\bibitem{Dynamical_exponent} The gauge action resulting from critical
boson fluctuations is well known for the relativistic case, given
by\cite{Kleinert} \bqa && S_{a} =
\int_{0}^{\beta}{d\tau}\int{d^2r}
\frac{N_{z}}{16}(\partial\times{a})\frac{1}{\sqrt{-\partial^{2}}}(\partial\times{a})
, \nonumber \eqa where $N_{z}$ is the flavor number of spinons. In
energy-momentum space the effective gauge action is \bqa && S_{a}
= \frac{N_{z}}{16}\sum_{q,\omega_{n}}\sqrt{{q^2}+\omega_{n}^{2}}
\Bigl(\delta_{ij} - \frac{q_{i}q_{j}}{q^2}\Bigr)a_{i}a_{j} ,
\nonumber \eqa where the time component $a_{\tau}$ can be ignored
since their fluctuations are gapped, as mentioned in the text.
Thus, the total gauge action at the quantum critical point is
given by \bqa && S_{a} =
\frac{1}{2}\sum_{q,\omega_{n}}\Bigl(\frac{q^2}{g} +
\frac{E_{\eta}}{v_{\eta}}\frac{|\omega_{n}|}{q} +
\frac{N_{z}}{8}\sqrt{{q^2}+\omega_{n}^{2}}\Bigr) \Bigl(\delta_{ij}
- \frac{q_{i}q_{j}}{q^2}\Bigr)a_{i}a_{j} .  \nonumber \eqa To
determine the dynamical exponent $z$, we find the dispersion
relation for gauge fluctuations from the pole of the above gauge
kernel, given by in real frequency \bqa && \frac{q^2}{g} - i
\frac{E_{\eta}}{v_{\eta}}\frac{ \omega }{q} +
\frac{N_{z}}{8}\sqrt{{q^2}-\omega^{2}} = 0 . \nonumber \eqa
Solving this equation, we obtain \bqa && \omega = \frac{- i
\frac{E_{\eta}}{v_{\eta}g}q^{3} \pm
\sqrt{-\Bigl(\frac{E_{\eta}}{v_{\eta}g}\Bigr)^{2}q^{6} -
\Bigl[\Bigl(\frac{E_{\eta}}{v_{\eta}}\Bigr)^{2} -
\Bigl(\frac{N_{z}}{8}\Bigr)^{2}{q^2}\Bigr]\Bigl[\Bigl(\frac{N_{z}}{8}\Bigr)^{2}q^{4}
- \frac{q^{6}}{g^{2}}\Bigr]
}}{\Bigl(\frac{E_{\eta}}{v_{\eta}}\Bigr)^{2} -
\Bigl(\frac{N_{z}}{8}\Bigr)^{2}{q^2}} . \nonumber \eqa In the long
wave length limit ($q \rightarrow 0$) the dispersion becomes \bqa
&& \omega \approx \pm i
\frac{1}{\Bigl(\frac{E_{\eta}}{v_{\eta}}\Bigr)}\frac{N_{z}}{8}q^{2}
. \nonumber \eqa This implies that the dynamical exponent is $z =
2$ at the quantum critical point. The corresponding gauge action
is that in Eq. (40).
\bibitem{Ioffe-Larkin} One may expect that the response function such
as the conductivity is given by the Ioffe-Larkin combination rule
[L. B. Ioffe and A. I. Larkin, Phys. Rev. B {\bf 39}, 8988
(1989)]. However, this is not true owing to the presence of
$\sigma = \pm$ in the gauge coupling. Considering the conductivity
for example, one finds that it is solely given by the fermion
conductivity. Expanding the effective action to the gaussian order
for the U(1) gauge field $a_{\mu}$, the partition function is
given in a highly schematic form \bqa && Z = \int{Da}\exp\Bigl[-
\frac{1}{2}\frac{\delta^{2}S_{z}}{\delta{a}^{2}}{a}^{2} -
\frac{1}{2}\Bigl(\frac{\delta^{2}S_{\psi}}{\delta{a}^{2}}{a}^{2} +
2\frac{\delta^{2}S_{\psi}}{\delta{a}\delta{A}}{a}{A} +
\frac{\delta^{2}S_{\psi}}{\delta{A}^{2}}{A}^{2} \Bigr) \Bigr] \nn
&& = Z_{0}\exp\Bigl[ -
\frac{1}{2}\frac{\delta^{2}S_{\psi}}{\delta{A}^{2}}{A}^{2} +
\frac{1}{2}\frac{\frac{\delta^{2}S_{\psi}}{\delta{a}\delta{A}}}
{\frac{\delta^{2}S_{z}}{\delta{a}^{2}}+\frac{\delta^{2}S_{\psi}}{\delta{a}^{2}}}A^{2}\Bigr]
, \nonumber \eqa where $S_{z}$ ($S_{\psi}$) is the boson (fermion)
action. Differentiating the above effective action by an
electromagnetic field $A_{\mu}$ twice, one can obtain the
conductivity expression. The presence of $\sigma = \pm$ results in
${\delta^{2}S_{\psi}}/{\delta{a}\delta{A}} = 0$ because
correlations between the charge current and the spin current are
zero. As a result, the conductivity is given by the fermion
contribution only.
\end{thebibliography}
\end{document}